\documentstyle[prb,aps,epsfig,floats,amsfonts]{revtex}

\begin{document}

\twocolumn \psfull \draft

\wideabs{
\title{Quantum transport through mesoscopic disordered interfaces, junctions, and multilayers}
\author{Branislav K. Nikoli\' c$^*$}
\address{Department of Physics and Astronomy, SUNY at Stony Brook,
Stony Brook, NY 11794-3800}

\maketitle

\begin{abstract}
The study explores perpendicular transport through macroscopically
inhomogeneous three-dimensional disordered conductors using
mesoscopic methods (real-space Green function technique in a
two-probe measuring geometry). The nanoscale samples (containing
$\sim1000$ atoms) are modeled by a tight-binding Hamiltonian on a
simple cubic lattice where disorder is introduced in the on-site
potential energy. I compute the transport properties of:
disordered metallic junctions formed by concatenating two
homogenous samples with different kinds of microscopic disorder,
a single strongly disordered interface, and multilayers composed
of such interfaces and homogeneous layers characterized by
different strength of the same type of microscopic disorder. This
allows us to: contrast resistor model (semiclassical) approach
with fully quantum description of dirty mesoscopic multilayers;
study the transmission properties of dirty interfaces (where Schep-Bauer
distribution of transmission eigenvalues is confirmed for single
interface, as well as for the stack of such interfaces that is
thinner than the localization length); and elucidate the effect of
coupling to ideal leads (``measuring apparatus'') on the conductance
of both bulk conductors and dirty interfaces  When multilayer contains
a ballistic layer in between two interfaces, its disorder-averaged
conductance oscillates as a function of Fermi energy. I also address
some fundamental issues in quantum transport theory---the relationship
between Kubo formula in exact state representation and ``mesoscopic Kubo
formula'' (which gives the zero-temperature conductance of a
finite-size sample attached to two semi-infinite ideal leads) is
thoroughly reexamined by comparing their answers for both the
junctions and homogeneous samples.

\end{abstract}

\pacs{PACS numbers:  73.40.-c, 72.10.Fk, 73.63.-b, 05.60.Gg}}

\narrowtext

\section{Introduction}

The experimental discovery~\cite{gmr1,gmr2} of a giant
magnetoresistance (GMR) phenomenon has revived interest in the
transport properties of macroscopically inhomogeneous conductors,
such as metallic junctions~\cite{dugaev} or
multilayers.~\cite{gijs} Furthermore, unusual systems for
traditional transport theory, like  single dirty
interface~\cite{schep-bauer} which are ubiquitous elements in
such circuits, have entailed the introduction of new concepts to
replace usual quantities (e.g., mean free path $\ell$) used to
describe transport in bulk samples. Although some of these
problems were formulated long ago within the (semiclassical)
transport theory,~\cite{fan,fuchs,sondh} new attacks have
employed all (quantum and semiclassical) transport  formalisms
developed thus far revealing that such problems are by no means
resolved.~\cite{gijs,butler} In particular, the reexamination of
various fundamental issues in the transport theory has been
brought about by the experimental and theoretical advances in
mesoscopic physics.~\cite{lesh} Thus, the Landauer-B\" uttiker
scattering formalism~\cite{lb} has been frequently invoked to
study transport in both non-magnetic~\cite{bauermetal} and
magnetic multilayered conductors.~\cite{bauerprl,asano}
Obviously, the thorough understanding of transport properties due
to purely {\it multilayer + disorder} effects is prerequisite for
the analysis of more complicated phenomena in inhomogeneous
structures.

Besides providing the means to compute the (quantum) conductance
of finite-size samples, mesoscopic methods give additional
physical insight by delineating transmission properties of the
sample (note that one can technically treat transport in
macroscopic samples without really using such fully
quantum-mechanical transport theories~\cite{gijs}). The
finite-size of mesoscopic systems play an important role in
determining the conductance through scattering approach, but no
further limitations exist---the exactness of results obtained in this
guise is heavily exploited in throughout the paper. Practical
realization of this program appears in different incarnations, i.e.,
different Landauer-type~\cite{caroli} or Kubo~\cite{kubo-nikolic} formulas
for a finite-size phase-coherent sample attached to semi-infinite
ideal (i.e., disorder-free) leads. These  prescriptions are usually
made computationally efficient by combining them with some Green function
technique.~\cite{lambert,datta}

Here I employ  mesoscopic quantum transport methods to calculate
the conductance of disordered samples which are macroscopically
inhomogeneous, i.e., composed of different homogeneous conductors
(``layers'') joined through some interfaces (``monoatomic
layers''). In homogeneous conductors, whose properties are
well-studied throughout the history of localization
theory,~\cite{kramer} impurities generate only microscopic
inhomogeneity on the scale $\sim \lambda_F$ (Fermi wavelength).
Our goal is twofold: ({\bf 1}) Most mesoscopic studies have been
focused on the bulk homogeneous conductors~\cite{datta} in the
weak scattering transport regime. Only recently the systems like
metallic multilayers,~\cite{bauermetal} or single dirty
interfaces~\cite{schep-bauer,yehuda} have been tackled in this
guise. By employing nonperturbative numerical methods we can
access strongly disordered junctions, single strongly disordered
interface (when stacked together to form a bulk conductor our
interfaces would form an Anderson insulator), and multilayers
composed of such interfaces and bulk diffusive or ballistic
mesoscopic conductors. A system is called mesoscopic if its size
$L$ is smaller than the dephasing length $L_\phi$, which is a
typical distance for electron to travel without loosing its phase
coherence ($L_\phi \lesssim 1 \,\mu$m in current low-temperature
experiments), and is therefore determined by decoherence
processes caused by the coupling to environment either through
inelastic scattering (electron-electron and electron-phonon) or
just be the change of the environment quantum state (e.g.,
spin-flip scattering from a magnetic impurity). The term metallic
implies that conductance $G$ of a bulk homogeneous sample is much
larger than the conductance quantum $G_Q=2e^2/h$. For interfaces
one needs a different nomenclature: they are termed
``dirty''~\cite{schep-bauer} if their conductance $G/G_Q$ is much
smaller than the number of conducting channels $N_{\rm ch}$
($N_{\rm ch} \sim k_F^2 S$ in three-dimensions, with
$k_F=2\pi/\lambda_F$, and $S$ is the cross section of a sample).
Lacking a better language, I denote the multilayers studied here
as ``dirty metallic'', meaning that scattering is due to a random
potential and their conductance is $G_M \ll G_Q N_{\rm ch}$.
Nevertheless, layer components are chosen to be metallic $G \gg
G_Q$, and are well-described, in the diffusive limit $L \gg
\ell$, by semiclassical transport theory.~\cite{nikolic_qrho} In
terms of material parameters, the resistivity of a bulk
homogeneous material, from which nanoscopic layers are cut out,
is few $100\, \mu \Omega$cm, which is typical of dirty transition
metal alloys. Thus, the possible small conductance of layers is
not necessarily caused by approaching the
localization-delocalization transition~\cite{janssen} upon
increasing disorder, as is usual in the homogeneous bulk samples.
The conductors are modeled by a tight-binding Hamiltonian with
on-site potential disorder. This corresponds to a model of free
electrons (understood here as Landau quasiparticles with
parameters renormalized by both band structure effects and Fermi
liquid interaction)  with random point scatterers,~\cite{butler}
which is used frequently in studies of similar systems. While
isotropic scattering sets semiclassical vertex corrections to
zero (which are determined by ladder diagrams~\cite{rammer}
generating difference between momentum relaxation time and
elastic mean free time, or scattering-in term in the Boltzmann
theory), it does not eliminate the higher order quantum
interference vertex corrections. These terms are nonlocal on the
mesoscopic length scale $L_\phi$, and therefore
invalidate~\cite{gijs} the concept of local position dependent
conductivity $\sigma(x)$ as usual way of describing the
multilayered structures (in semiclassical
approximation).~\cite{butler} Since I exploit here exact
techniques, all quantum localization effects which are not
necessarily small in dirty systems,~\cite{butler,nikolic_qrho}
are included from the onset. All three types of samples are
studied for electron transport perpendicular to the layers (or
interfaces), which is the so-called current perpendicular to
plane (CPP) geometry.~\cite{gijs} Once the disorder-averaged
resistance of the multilayer is computed, we can compare it to
the resistance given by the resistor model,~\cite{levy} (i.e., a
sum of the bulk layers and interface resistances connected like
classical Ohmic resistors in a series). ({\bf 2}) I investigate
some fundamental issues in the quantum transport theory using
dirty metallic junctions from above as a testing ground,  as well
as homogeneous disordered samples as a reference. That is, I
compare the transport properties computed from the Kubo formula
in exact single particle state representation (which was widely
used~\cite{krey} in the ``premesoscopic'' era~\cite{leeandrama}
of the Anderson localization theory) and ``mesoscopic Kubo
formula'' for the open finite-size system attached to two ideal
leads.~\cite{kubo-nikolic} In the former case the system is
closed and the eigenproblem of the Hamiltonian is solved exactly
by numerical diagonalization, while in the latter case energy
levels of a sample are broadened by the coupling to the leads,
and I use real-space Green functions for such open sample plugged
into the Kubo formula~\cite{kubo-nikolic} (which is then
equivalent to the Landauer two-probe
formula~\cite{baranger-kubo}) to get its conductance. Also, the
influence of the leads (``measuring apparatus'') and the
lead-sample coupling on the conductance (which is akin to the
problems encountered in quantum measurement
theory~\cite{efetovbook}) is explicitly quantified.

The paper is organized as follows. The first Sec.~\ref{sec:kubo}
deals with some general remarks on the Kubo linear response
formalism and current conservation. This should serve as a
guidance for proper application of Kubo formulas on the
finite-size systems. In Sec.~\ref{sec:metaljunc}, the Kubo
formula in exact state representation is compared to the Kubo
formula for the finite-size sample by evaluating them for dirty
metallic junction, as well as for homogeneous samples.
Section~\ref{sec:interface} presents the results on the
transmission properties, and conductance derived from them, of a
single dirty interface as well as thin layers composed of few of
such interfaces. The calculations are completed in
Sec.~\ref{sec:multilayer} by examining different types of
multilayers, containing disordered and/or ballistic layers
concatenated through interfaces from Sec.~\ref{sec:interface}.
Section~\ref{sec:conclusion} provides a summary of the technical,
emphasizing physical insights gained from them.

\section{Kubo formulas and current conservation}
\label{sec:kubo}

The basic global transport property, for small applied voltage,
is linear conductance (or equivalent resistance $R=1/G$).
The conductance $G$ is defined by the Ohm's law
\begin{equation}\label{eq:ohm}
  I=G \, V,
\end{equation}
relating the current $I$ to the voltage drop $V$. Since this is a
plausible relation for a linear transport regime, more information
is contained in the local form of the Ohm's law
\begin{equation}\label{eq:locohm}
  {\bf j}(\bf{r},\omega) = \int \! d {\bf r}^\prime \, \mbox{\b{$\sigma$}}
  ({\bf r},{\bf r}^\prime;\omega) \cdot {\bf E}({\bf r}^\prime,\omega).
\end{equation}
In what follows, the focus will be on the DC transport $\omega
\rightarrow 0$. Although it is possible to treat ${\bf E}({\bf
r})$ as an externally applied electric field and then include the
effects of Coulomb interaction between electrons as a
contribution to the vertex correction,~\cite{gijs} the usual
approach is to use the total (local and inhomogeneous) electric
field ${\bf E}({\bf r}) \equiv {\bf E}_{\rm loc}({\bf r})=-\nabla
\mu({\bf r})/e$, which is the sum of external field plus the
field due to the charge redistribution from the system
response.~\cite{butlerlrt} The electrons are then treated as
independent quasiparticles. The electrochemical potential $\mu$ in
non-equilibrium situations (like transport) is not a well-defined
quantity, and can serve only e.g., to parameterize the carrier
population.~\cite{payne} The relation~(\ref{eq:locohm}) defines
the meaning of the nonlocal conductivity tensor (NLCT)
$\mbox{\b{$\sigma$}} ({\bf r},{\bf r}^\prime)$ as the fundamental
microscopic quantity in the linear response theory. This quantity
gives the current response at ${\bf r}$ due to an electric field
at ${\bf r}^\prime$. The requirement of current conservation in
the DC transport
\begin{equation}\label{eq:divj}
  \nabla \cdot {\bf j}({\bf r}) = 0,
\end{equation}
coupled to Eq.~(\ref{eq:locohm}) and together with appropriate
boundary conditions at the infinity or possible
interfaces,~\cite{nazarov} makes it possible to find ${\bf E}({\bf r})$.
The conductance (dissipated power by the voltage $V$ squared) is
then given by
\begin{eqnarray}\label{eq:volumecond}
  G & = & \frac{1}{V^2} \int\limits_{\Omega} d{\bf r} \, {\bf E}({\bf r})
  \cdot {\bf j}({\bf r})  \nonumber \\
  & = & \frac{1}{V^2} \int\limits_{\Omega}  d{\bf r} \, d{\bf r}' \, {\bf E}({\bf r})
  \cdot \mbox{\b{$\sigma$}} ({\bf r},{\bf r}') \cdot {\bf E}({\bf r}^\prime),
\end{eqnarray}
where $\Omega$ is the sample volume, and at first sight requires
the knowledge of local electric fields within the sample.
However, it was shown in the course of recent reexamination of
transport theory,~\cite{kane} driven by the problems of
mesoscopics, that current conservation imposes stringent
requirements on the form of NLCT
\begin{equation}\label{eq:divergeless2}
  \nabla \cdot \mbox{\b{$\sigma$}} ({\bf r},{\bf r}')=\mbox{\b{$\sigma$}}
  ({\bf r},{\bf r}') \cdot \nabla'=0,
\end{equation}
where absence of magnetic field is assumed to get this special
case of a more general theorem.~\cite{baranger-kubo} This allows
us to use arbitrary electric field factors in Eq.~(\ref{eq:volumecond}),
including ${\bf E}({\bf r}) \neq {\bf E}({\bf r}^\prime)$ [a
homogenous field $E=V/L$ is usually implied in the textbook
literature~\cite{rammer}].

When a finite-size sample is attached to two semi-infinite ideal
leads, the condition~(\ref{eq:divergeless2}) is sufficient to show
that~\cite{baranger-kubo}
\begin{equation}\label{eq:condsurf}
  G= - \int\limits_{S_1} \!  \int\limits_{S_2} d{\bf S}_1 \cdot \mbox{\b{$\sigma$}}
  ({\bf r},{\bf r}') \cdot d{\bf S}_2,
\end{equation}
by using the divergence theorem to push the integration from the
bulk onto the boundary surface~\cite{kubosubtle} going through
the leads and around the disordered sample (the integration over
this insulating boundary obviously gives zero contribution
because no current flows out of it). The surface integration in
the two-probe conductance formula~(\ref{eq:condsurf}) is over
surfaces $S_1$ and $S_2$ separating the leads from the disordered
sample. The vectors $d{\bf S}_1$ and $d{\bf S}_2$ are normal  to
the cross sections of the leads, and are directed outwards from
the region encompassed by the overall surface (composed of $S_1$,
$S_2$ and insulating boundaries of the sample). It is assumed
that voltage in one of the leads is zero, e.g., $\mu_L=0$ and
$\mu_R=e V$. The meaning of current conservation and
expression~(\ref{eq:condsurf}) is quite transparent---current $I$
at a given cross section depends only on the voltages in the leads (in
experiment one either fixes this voltages by a voltage source, or
fixes the current by a current source) and not on the precise
electric field configurations. The formula~(\ref{eq:condsurf})
can be generalized~\cite{baranger-kubo} to arbitrary multi-probe
geometry, while the volume-averaged
conductance~(\ref{eq:volumecond}) is meaningful only for the
two-probe measurement. The expression~(\ref{eq:condsurf}) is valid
even in the presence of interactions, where many-body effects can
be introduced using Kubo formalism to get $\mbox{\b{$\sigma$}}
({\bf r},{\bf r}')$ microscopically. However, this route is
tractable and useful especially in the case of noninteracting
quasiparticle systems. When a sample is attached to ideal leads
(see below), this established rigorous equivalence~\cite{baranger-kubo}
between two different linear response formulations---Kubo and
Landauer-B\"uttiker~\cite{lb} (to work out the proof, $S_1$ and $S_2$ should be
placed far enough into the leads so that all evanescent modes
from the sample have died out and do not contribute to the
conductance~\cite{gijs}). In the scattering approach to transport,
pioneered by Landauer through heuristic and subtle arguments, the
conductance of a noninteracting system is then expressed in terms
of the transmissions probabilities between different quantized
transverse propagating modes defined by the leads as asymptotic scattering
states [see Eq.~(\ref{eq:ttlandauer})].

Although Boltzmann formalism can provide semiclassical expression for
NLCT (which is nonlocal on the scale of the sample size because of
the classical requirement of current conservation~\cite{palee1}), the standard
quantum route to it is the Kubo linear response theory (KLRT). Linear
response theory gives a full quantum description of transport
(i.e., it includes quantum interference effects in the motion of
electrons~\cite{seemtobe}) in non-equilibrium systems that are still
close enough to equilibrium for vanishingly small applied external
electric field.~\cite{linearity} Therefore,
the formulas above involve the equilibrium expectation values of
the corresponding quantum-mechanical operators, in accordance
with fluctuation-dissipation theorem underlying KLRT [e.g., it is
assumed that ${\bf j}(\bf{r})$ is the expectation value of the
current density operator in the quantum formalism]. While the
current is response to the total electric field ({\em external + induced}),
the NLCT is obtained as a response to the external field only~\cite{stonebook,orifice}
because the current induced by the external field is already linear
in the field. Therefore, in the linear response one does not need
the corrections due to induced non-equilibrium charges (electron-electron
interaction is taken in the equilibrium, e.g. through renormalized parameters
of the Landau quasiparticles and the self-consistent screening effect on
the impurity scattering).

Since Kubo NLCT is not experimentally measurable quantity,
macroscopic conductivity $\sigma$ is obtained by volume averaging
NLCT through Eq.~(\ref{eq:volumecond}) where $\sigma = G \, L/S$
for a cubic sample of length $L$ and cross section $S$ (limit
$\Omega=LS \rightarrow \infty$ is assumed, while keeping the
impurity concentration finite).  This usually implies a
homogeneous ${\bf E}({\bf r})$ factors, which is justified by the
fact that Kubo expression for NLCT is divergenceless (as can be
easily proven~\cite{kane} from its current-current correlation
form,~\cite{rammer}) i.e., it satisfies
Eq.~(\ref{eq:divergeless2}). Thus, the volume-averaged
conductivity $\sigma$ relates the spatially averaged current
${\bf j}=\int d{\bf r} \, {\bf j}({\bf r})/\Omega$ to the
spatially-averaged electric field,
\begin{equation}\label{eq:macrosigma}
{\bf j}=\sigma {\bf E}.
\end{equation}
For a noninteracting system of fermions this leads to a Kubo
formula in exact single-particle state representation (KFESR)
\begin{eqnarray}\label{eq:kuboesr}
  \sigma_{xx} & = & \frac{2 \pi \hbar e^2}{\Omega} \sum_{\alpha, \, \alpha^\prime}
   |\langle \alpha | \hat{v}_x | \alpha^\prime \rangle|^2
   \nonumber \\
   && \times \delta(E_\alpha - E_F)
   \delta(E_{\alpha^\prime}-E_F),
\end{eqnarray}
where the factor of two for spin degeneracy will be  explicit in
all formulas. The delta functions in Eq.~(\ref{eq:kuboesr})
emphasize that conductivity is a Fermi surface ($E_F$ is the
Fermi energy) property at low temperatures [$T \rightarrow 0
\Rightarrow -\partial f(E)/\partial E \simeq \delta(E_F-E)$,
$f(E)$ being the Fermi-Dirac distribution function]. Here
$\hat{v}_x$ is the $x$-component of the velocity operator and
$|\alpha \rangle$ are eigenstates of a single-particle
Hamiltonian, $\hat{H}|\alpha \rangle =E_\alpha|\alpha \rangle$.
The velocity operator $\hat{\bf v}$ is determined through the
equation of motion for the position operator $\hat{\bf r}$
\begin{equation}\label{eq:velocityop}
  i \hbar \hat{{\bf v}}= i \hbar \frac{d\hat{{\bf r}}}{dt}=[\hat{{\bf
  r}},\hat{H}],
\end{equation}
where $\hat{H}$ is the Hamiltonian before the application of
the external electric field (in the spirit of FDT).

In the general case, conductivity is a tensor, but since symmetries
are restored after disorder-averaging, one can use
$\sigma=(\sigma_{xx}+\sigma_{yy}+\sigma_{zz})/3$ as the scalar conductivity.
This is valid only in the case of homogenously disordered sample.
For example, in our metal junction or multilayers $\sigma_{xx}$
is different from $\sigma_{yy}$ and $\sigma_{zz}$. To get the
conductivity (as an intensive quantity) in KLRT, the
thermodynamic limit $\Omega$ is always understood (no stationary
regime can be reached in a system which in neither infinite nor
coupled to some thermostat). In the disorder electron physics
this also bypasses the ambiguity of conductivity which
scales~\cite{gang4} with the length of the system (although
scaling is unimportant~\cite{janssen} in the metallic regime $G
\gg G_Q$). Nonetheless, the computation of transport properties
from exact single particle eigenstates, obtained by the numerical
diagonalization of Hamiltonian of a finite-size system, has been
frequently employed in the development of disordered electron
physics.~\cite{krey} In fact, it is still the standard method for
computing the conductivity of many-body systems by diagonalizing small
clusters of some lattice fermion Hamiltonian.~\cite{rici}
However, direct application of the formula~(\ref{eq:kuboesr})
leads to a trouble since eigenvalues are discrete when the sample
is finite and isolated. Strictly speaking, the conductivity is
then a sum of delta functions, and to obtain finite conductivity
they have to be broadened into functions having a finite width
larger than the level spacing. Thus, there are two numerical
tricks which can be used to ``circumvent'' this problem: (i) one
can start from the Kubo formula for the frequency dependent
conductivity
\begin{eqnarray}\label{eq:condkuboomega}
   \sigma_{xx}(\omega) & = & \frac{2 \pi \hbar e^2}{\Omega} \sum_{\alpha, \, \alpha^\prime}
   |\langle \alpha | \hat{v}_x | \alpha^\prime \rangle|^2
   \nonumber \\
   && \times \frac{f(E_\alpha)-f(E_{\alpha^\prime})}{\hbar \omega}
   \delta(E_\alpha - E_{\alpha^\prime} - \hbar \omega),
\end{eqnarray}
average the result over finite $\omega$ values, and finally
extrapolate~\cite{allenglass} to the static limit $\omega
\rightarrow 0$; (ii) The delta functions in Eq.~(\ref{eq:kuboesr})
can be broadened into a Lorentzian
\begin{equation}\label{eq:lorentz}
  \delta(x) \rightarrow \bar{\delta}(x)=\frac{1}{\pi}
  \frac{(\eta/2)^2}{x^2+(\eta/2)^2},
\end{equation}
where $\eta$ is the full width at half maximum of the Lorentzian.
I find that both methods produce similar results. The
calculations presented below use the broadened delta function
$\bar{\delta}(x)$ in the formula for static Kubo conductivity.
\begin{figure}
\centerline{\psfig{file=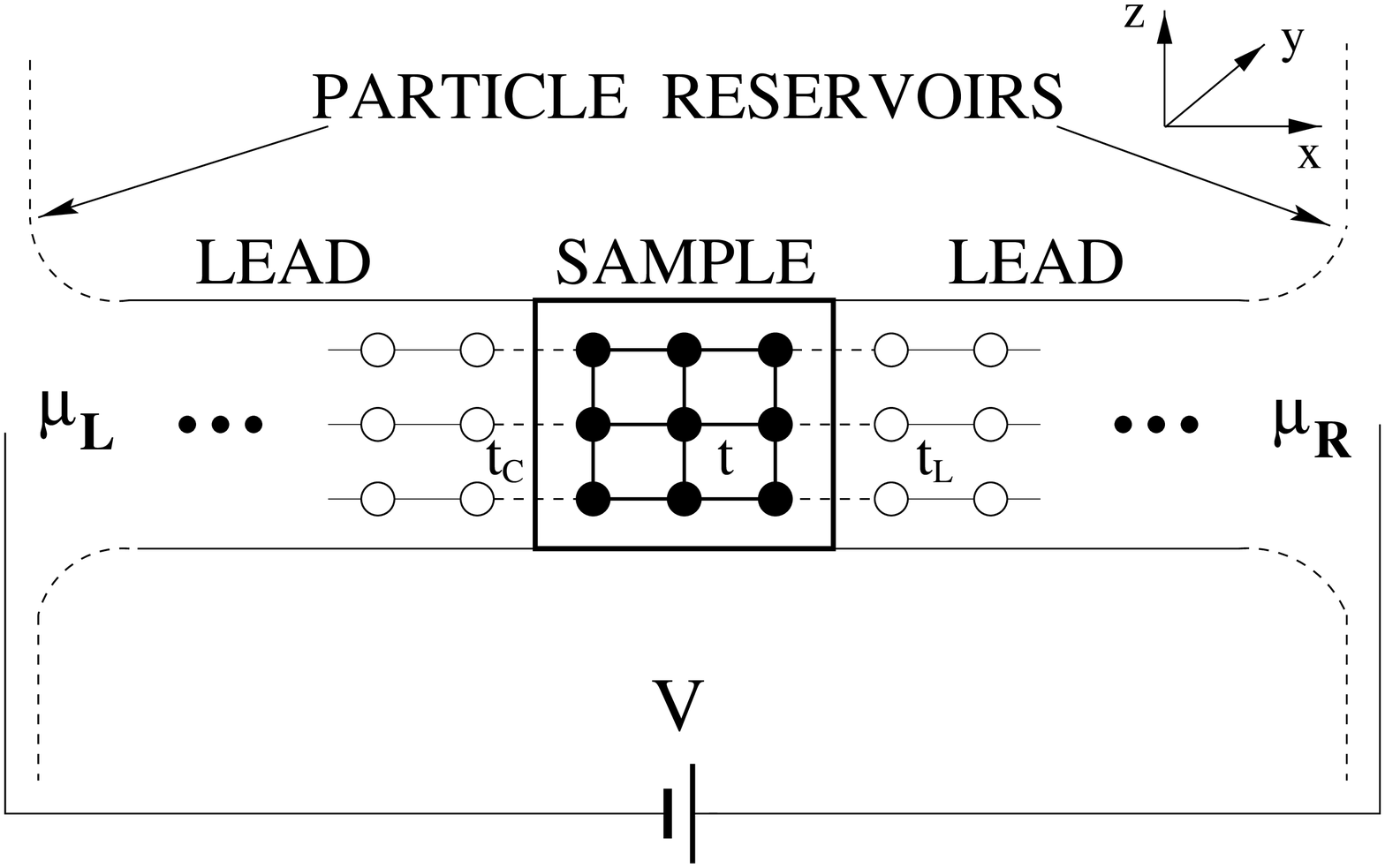,height=2.3in,width=3.0in,angle=0}
} \vspace{0.2in} \caption{A two-dimensional version of the actual
3D model used here for a two-probe measuring geometry. Each site hosts a
single $s$-orbital which hops to six (or fewer for surface atoms)
nearest neighbors. The hopping matrix element is $t$ (within the
sample), $t_{\rm L}$ (within the leads), and $t_{\rm C}$
(coupling of the sample to the leads). The leads are
semi-infinite and connected smoothly at $\pm \infty$ to
macroscopic reservoirs biased by the chemical potential difference
$\mu_L-\mu_R=eV$.} \label{fig:setup}
\end{figure}

The Green operator of a noninteracting Hamiltonian (with
specified boundary conditions),
\begin{equation}\label{eq:greeneta}
  \hat{G}^{r,a}=[E-\hat{H} \pm i \eta]^{-1},
\end{equation}
contains the same information encoded in the single-particle wave
function. Therefore, using the expansion of single-particle Green
operator $\hat{G}^{r,a}=\sum_\alpha |\alpha \rangle \langle
\alpha|/(E-E_\alpha \pm i\eta)$ in terms of exact eigenstates,
the Kubo formula for the macroscopic (volume-averaged) longitudinal
conductivity can be recast into the following
expression~\cite{kubo_volume}
\begin{mathletters}
\label{eq:greenkuboold}
\begin{eqnarray}
 \sigma_{xx} & = &  \frac{2\pi e^2 \hbar}{\Omega} \, \text{Tr} \left
   [\hat{v}_x\, \delta(E-\hat{H}) \, \hat{v}_x  \, \delta(E-\hat{H})  \right
  ], \\
 \delta(E-\hat{H}) & = & -\frac{1}{\pi}\text {Im} \, \hat{G}
  =  \frac{1}{2i} (\hat{G}^r - \hat{G}^a).
\end{eqnarray}
\end{mathletters}
When employing this formula one faces the same problem as in the
KFESR---in order to define~\cite{economou} retarded ($r$) or
advanced ($a$) Green function a small parameter $\eta \rightarrow
0^+$ requires numerical handling~\cite{kirkpatrick} analogous to
the one~(\ref{eq:lorentz}) involved in the
KFESR~(\ref{eq:kuboesr}). However, it is possible to derive
another, ``mesoscopic Kubo formula'', for the finite-size sample
attached to two semi-infinite clean leads.~\cite{baranger-kubo}
This formula represent a fully quantum-mechanical expression
for the zero-temperature conductance
\begin{equation}\label{eq:greenkubo}
   G_{xx}  =  \frac{4e^2}{h} \frac{1}{L_x^2} \, {\rm Tr} \left (\hbar \hat{v}_x
 {\rm Im} \,  \hat{G} \, \hbar \hat{v}_x {\rm Im} \, \hat{G} \right
 ),
\end{equation}
which, being measurable, is the only meaningful quantity to
discuss in mesoscopics  (the alternative is to introduce
conductivity as the NLCT). Namely, in quantum-coherent samples
($L > L_\phi$) local description of transport in terms of
conductivity breaks down because of the nonlocal correction
(e.g., weak localization~\cite{wl}) induced on the scale of
$L_\phi \gg \ell$, which is much larger than the elastic mean free
path. Thus, mesoscopic samples have to be treated as a single
coherent unit (``giant molecule''), so that its conductance
cannot be interpreted as a combination $G=\sigma A/L$ of the
conductances of its parts (i.e., such description becomes
applicable only at high enough temperatures). Although the
formula~(\ref{eq:greenkubo}) apparently looks like
Eq.~(\ref{eq:greenkuboold}), the crucial difference is that the
following Green operator is plugged there instead
\begin{equation}\label{eq:greenleads}
   \hat{G}^{r,a}=[E-\hat{H} - \hat{\Sigma}^{r,a}]^{-1}.
\end{equation}
This change is not innocuous: $\hat{G}^{r}$
($\hat{G}^{a}=[\hat{G}^{r}]^{\dagger}$,
$\hat{\Sigma}^{a}=[\hat{\Sigma}^{r}]^{\dagger}$) has acquired a
self-energy term from the ``interaction'' with the
leads.~\cite{datta} This looks like the self-energy term in the
many-body Green functions, but this one is exactly calculable, as
shown below. Thus, the leads enforce new boundary conditions.
Since $\hat{\Sigma}^r(E)$ contains an imaginary part when energy
$E$ belongs to the band of a lead, no small parameter $\eta$ is
needed to define the retarded or advanced Green operator.

The lattice model of a two-probe measuring geometry, to be used
in the subsequent sections for evaluation of
Eq.~(\ref{eq:greenkubo}), is shown in Fig.~\ref{fig:setup}. The
three-dimensional (3D) nanocrystal (``sample'') is placed between
two ideal (disorder-free) semi-infinite ``leads'' which are
connected smoothly to macroscopic reservoirs at infinity. The
electrochemical potential difference $e V=\mu_L-\mu_R$, which
drives the current,~\cite{landimry} is measured between the
reservoirs. It is assumed that reservoirs inject thermalized
electrons at electrochemical potential $\mu_L$ (from the left) or
$\mu_R$ (from the right) into the system. All inelastic
scattering occur in the reservoirs,  which therefore ensure
steady state transport in the central region. Semi-infinite
leads~\cite{fisher} are a convenient method to take into account
electrons entering of leaving the phase-coherent sample (i.e.,
effective Hamiltonian of an open system, $\hat{H}+
\hat{\Sigma}^r$ in Eq.~(\ref{eq:greenleads}), is non-Hermitian).
This makes it possible to bypass explicit modeling of
thermodynamics of perfect (i.e., unaffected by the flow of
current) macroscopic reservoirs which are introduced
heuristically in the scattering approach (for a different
interpretation of the role of perfectly conducting leads in the
derivation of Kubo formula for a finite-size system see
Ref.~\onlinecite{butlerlrt}). The leads have the same cross
section as the sample, which eliminates scattering induced by the
wide to narrow geometry at the sample-lead
interface.~\cite{szafer} The whole system is described by the
following Hamiltonian with nearest neighbor hopping integrals
$t_{\bf mn}$
\begin{equation}\label{eq:tbh}
  \hat{H} = \sum_{\bf m} \varepsilon_{\bf m}|{\bf m} \rangle \langle {\bf m}|
  +  \sum_{\langle {\bf m},{\bf n} \rangle} t_{\bf mn}
  |{\bf m} \rangle \langle {\bf n}|.
\end{equation}
where $\langle {\bf r}|{\bf m}\rangle$ is the $s$-orbital $\psi({\bf r}-{\bf m})$
located on  site ${\bf m}$. The site representation, defined by the basis states
$|{\bf m}\rangle$, can be interpreted either as a tight-binding description of
electronic states or as discretization of corresponding single-particle
Schr\" odinger equation. The sample is the central section with $N
\times N_y \times N_z$ sites. The hopping in the sample set the
unit of energy $t_{\bf mn}=t$. The disorder will be introduced by
taking $\varepsilon_{\bf m}$ to be a random variable. The leads
are clean ($\varepsilon_{\bf m}=0$) but, in general, have
different hopping integrals $t_{\bf mn}=t_{\text L} \neq t$.
Finally, the hopping which couples sample to the leads
is $t_{\bf mn}=t_{\text C}$. Hard wall boundary conditions are
set in the $\hat{y}$ and $\hat{z}$ directions. The different
hopping integrals introduced here are necessary when studying
disordered samples to get the conductance at Fermi energies
throughout the whole band extended (compared to the clean case)
by disorder, i.e., in such calculations one has to set $t_{\text L} >
t$.

In site representation, Green operator $\hat{G}^{r,a}$ becomes a
Green function matrix $\hat{G}^{r,a}({\bf n},{\bf m})=\langle
{\bf n} | \hat{G}^{r,a} | {\bf m} \rangle$. The self-energy
$\hat{\Sigma}^{r}=\hat{\Sigma}_L^{r}+\hat{\Sigma}_R^{r}$
``measures'' the coupling of the sample to the leads. This causes
the broadening of initially discrete levels and, thus, the finite
lifetime of electron in the sample. Since electron which leaves
the sample does not return phase-coherently, the dephasing length
is by definition $L_\phi = L_x$. Each lead generates its
self-energy term which has nonzero matrix elements only on the
edge layers of the sample which are adjacent to the leads. They
are defined as
\begin{equation} \label{eq:surface}
\hat{\Sigma}^{r}_{L,R}({\bf n},{\bf m}) = t_{\text C}^2
\hat{g}_{L,R}^r({\bf n}_S,{\bf m}_S),
\end{equation}
with $\hat{g}_{L,R}^r({\bf n}_S,{\bf m}_S)$ being the surface
Green function of a bare semi-infinite lead between the sites
${\bf n}_S$ and ${\bf m}_S$ in the end atomic layer of the
lead~\cite{datta} (adjacent to the corresponding sites ${\bf n}$
and ${\bf m}$ inside the conductor). I provide here explicit
expression for these self-energy terms in the most general case
when hopping integrals $t \neq t_{\text C} \neq t_{\text L}$ are
different in different parts of the setup in Fig.~\ref{fig:setup}
\begin{eqnarray}\label{eq:sigmanm}
  \hat{\Sigma}^{r}_{L,R}({\bf n},{\bf m}) & = & \frac{2}{N_y+1}
  \frac{2}{N_z+1} \sum_{k_y,k_z}
  \sin( k_y n_y a) \sin(k_z n_z a) \nonumber \\*
  & & \times \hat{\Sigma}^r (k_y,k_z) \sin( k_y m_y a) \sin(k_z
  m_z a).
\end{eqnarray}
This expression is obtained by expanding the surface Green
function $\hat{g}_{L,R}^r({\bf n}_S,{\bf m}_S)$ in terms of exact
eigenstates of the Hamiltonian of a semi-infinite lead, which satisfy
\begin{figure}
\centerline{\psfig{file=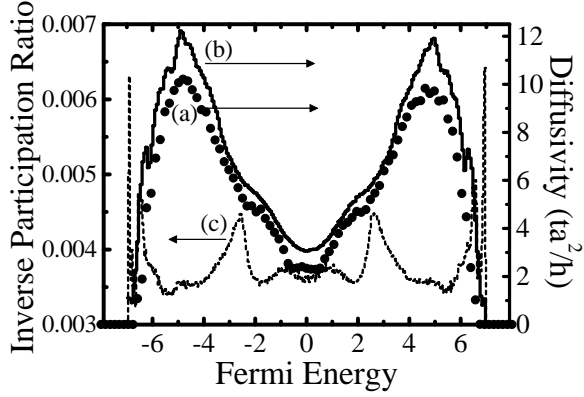,height=3.0in,angle=-90} }
\vspace{0.2in} \caption{Diffusivity $\langle D_x (E_F)
\rangle$ of a disordered binary alloy modeled by a tight-binding
Hamiltonian with quenched disorder ($\varepsilon_A=-2$ and
$\varepsilon_B=2$) on a lattice $18 \times 8 \times 10$: (a)
computed using the Kubo formula~(\ref{eq:greenkubo}) in terms of
the Green function for the sample with attached leads; (b)
computed from the Kubo formula in exact single-particle eigenstate
representation~(\ref{eq:statediff}), where the width of the
Lorentzian broadened delta function is $\eta=25 \Delta(E_F)$. Also
plotted (c) is the IPR, defined in Eq.~(\ref{eq:ipr}), which gives insight
(i.e., degree of localization) into the structure of eigenstates used to
evaluate the Kubo formula Eq.~(\ref{eq:statediff}). The disorder averaging
is performed over 50 samples.} \label{fig:diffbahomo}
\end{figure}
particular hard wall transverse boundary conditions. Here $({\bf
n},{\bf m})$ is the pair of sites on the surfaces inside the
sample which are adjacent to the leads $L$ or $R$, in accordance
with~(\ref{eq:surface}). Since electrons are confined in the $y$
and $z$ direction, electronic states in the lead have transverse
wave function which are labeled by the discrete quantum numbers.
They define subbands and corresponding conducting ``channels''
[i.e., transverse propagating modes as the product of transverse
wave function and Bloch wave in the $x$-direction, $\langle {\bf
m} |k_y,k_z \rangle \otimes |k_x \rangle$], which represent a
basis of scattering states for the scattering matrix of the
disordered region, as envisaged in the Landauer picture of quantum
transport. The self-energy $\hat{\Sigma}^r (k_y,k_z)$ of the
channel $(k_y,k_z)$ is given by
\begin{equation}\label{eq:sigmakk1}
  \hat{\Sigma}^{r}(k_y,k_z)=\frac{t_{\text C}^2}{2t_{\text L}^2} \left (E_{\Sigma}-i
  \sqrt{4t_{\text L}^2-E_{\Sigma}^2} \right ),
\end{equation}
for $|E_{\Sigma}|<2t_{\text L}$. I use the shorthand notation
$E_\Sigma=E-\varepsilon(k_y,k_z)$, where
$\varepsilon(k_y,k_z)=2t_{\text L} [ \cos(k_y a) + \cos(k_z a) ]$
is the energy of quantized transverse levels in the lead. In the
opposite case $|E_{\Sigma}|>2t_{\text L}$ we get
\begin{equation}\label{eq:sigmakk2}
  \hat{\Sigma}^{r}(k_y,k_z)=\frac{t_{\text C}^2}{2t_{\text L}^2} \left ( E_{\Sigma}-\text{sgn}
  \, E_{\Sigma} \sqrt{E_{\Sigma}^2-4t_{\text L}^2} \right ).
\end{equation}
When the level spacing of the subbands is much smaller than $E_F$,
the number of channels at $E_F$ is large and the finite-size model
can describe a metal. The possibility to evaluate exactly these
self-energies allows us to avoid the inversion of an infinite
matrix which would formally give the Green function for the whole
system~\cite{datta} in Fig.~\ref{fig:setup}. The real-space Green function
technique described here, pioneered by Caroli {\em et
al.}~\cite{caroli} long before official inception of mesoscopic
physics, treats an infinite system with a continuous spectrum by
evaluating only the Green function between the states inside the
sample. This offers an alternative to handling a finite-size
sample through periodic boundary conditions, whose discrete
spectrum prevents a direct evaluation of the Kubo
formulas~(\ref{eq:kuboesr}) and~(\ref{eq:greenkuboold}). In
general, the technical advantage of Green function techniques is
that they can be used even when well-defined asymptotic
conducting channels in the scattering approach to quantum transport
are hard to defined (e.g., this is the case when leads have complicated
shape,~\cite{yeyati} making it hard to get explicitly the exact asymptotic
eigenstates and their eigenspectrum).
\begin{figure}
\centerline{\psfig{file=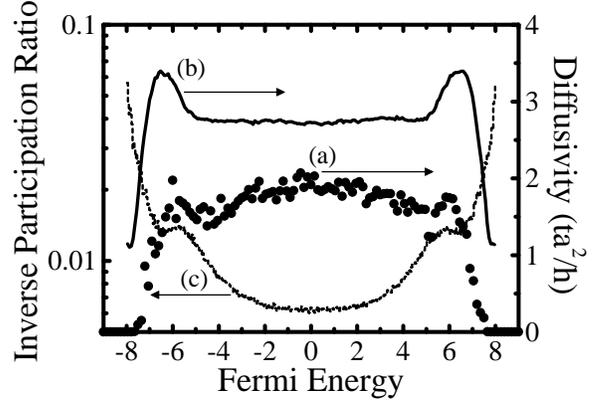,height=3.0in,angle=-90} }
\vspace{0.2in} \caption{The result of the same computation of
$\langle D_x (E_F) \rangle$ as in Fig.~\ref{fig:diffbahomo}, but
for a different type of disorder, diagonal one of the strength
$W=10$, introduced in the sample (the same labels apply to both
figures).} \label{fig:diffwhomo}
\end{figure}

The mesoscopic Kubo formula can be evaluated at
any (continuous) Fermi energy. At first sight, it seems like
that these leads to a much greater computational complexity than in
the case of KFESR because of the need to find inverse matrix in
Eq.~(\ref{eq:greenleads}) at each chosen energy, and then perform
the trace over the site states ${\rm Tr}(\ldots) = \sum_{\bf m}
\langle {\bf m}|(\ldots)|{\bf m}\rangle$. Namely, in the KFESR one
needs to diagonalize Hamiltonian only once and then use obtained
eigestates to compute matrix elements of velocity operator in the
eigenbasis. However, this is only an apparent difficulty once the
conservation of current is invoked. Since all Kubo formulas for
conductance stem from Eq.~(\ref{eq:volumecond}), one has to assume
(or choose) some electric field factors therein. Current
conservation, i.e., the fact that $I$ has to be the same on each
cross section, means that conductance is independent of this
choice. The minimal space for tracing, when system is described by
tight-binding Hamiltonian (TBH) in Eq.~(\ref{eq:tbh}), is obtained by
taking both factors ${\bf E}({\bf r})$ and ${\bf E}({\bf r}^\prime)$ to
be nonzero only on two adjacent planes of the lattice. Namely, the expectation value
of the velocity operator~(\ref{eq:velocityop}) in the site
representation is nonzero only between the states residing on
adjacent planes,
\begin{equation}\label{eq:tbhvelocity}
  \langle {\bf m} |\hat{v}_x| {\bf n} \rangle = \frac{i}{\hbar} t_{\bf
  mn} \left( m_x - n_x \right),
\end{equation}
for a TBH~(\ref{eq:tbh}) with nearest-neighbor hopping. The final
result is then to be divided by $a^2$ [instead of
$L_x^2$ in Eq.~(\ref{eq:greenkubo})].

Thus, only a block matrix $2N_z N_y \times 2N_z N_y$ of the Green function
for the whole sample has to be computed explicitly (e.g., those
elements which connect first two planes of the sample in
Fig.~\ref{fig:setup}). Because $[E-\hat{H}-\hat{\Sigma}^r]$ is a band-diagonal
matrix of a bandwidth $2N_yN_z+1$, one can shorten the
time needed to compute this block of the Green
function~(\ref{eq:greenleads}) by finding the LU decomposition of
a band-diagonal matrix, followed by forward-backward substitution
for each needed Green function element.~\cite{numrecipes} This
requires similar number of operations~\cite{verges_xi} as in the
more familiar recursive Green function technique.~\cite{fisher}

\section{Transport through dirty metal junctions}
\label{sec:metaljunc}

This section studies the static DC transport properties of a
metal junction composed of two disordered conductors with
different kind of disorder introduced on each side of the contact
interface. Both conductors are modeled as a disordered binary
alloy (i.e.,~composed of two types of atoms, $A$ and $B$) using
TBH of Eq.~(\ref{eq:tbh}). The quenched disorder is simulated by
taking the random on-site potential such that $\varepsilon_{\bf
m}$ is either $\varepsilon_A$ or $\varepsilon_B$ with equal
probability. Specifically, I take the lattice $18 \times 8 \times
10$ on each side of the junction and for the binary disorder:
$\varepsilon_A=-4$, $\varepsilon_B=0$ on the left; and
$\varepsilon_A=4$ and $\varepsilon_B=0$ on the right. Thus, the
junction has naturally a rough interface~\cite{asano} because of
the random positions of three different types of atoms around the
plane in the middle of the junction.

We commence with a description of the junction, as well as
homogeneous samples used as a reference, in terms of traditional
Kubo theory based on the KFESR~(\ref{eq:kuboesr}). Although
Ref.~\onlinecite{fisher} stated as one of the motivations for
undertaking the rigorous derivation of Landauer two-probe formula
from KLRT, that traditional use of Kubo formula~\cite{krey} was
numerically demanding, today's computers are much more powerful,
and it is of interest to compare this method to a modern~\cite{kubo-nikolic}
mesoscopic use of the Kubo formula. Kubo theory permits the discussion
of diffusivity $D_\alpha$ of an eigenstate $|\alpha \rangle$. This
quantity is extracted directly from the KFESR for Fermi gas of
noninteracting quasiparticles
\begin{equation}\label{eq:kubo-diffusivity}
    \sigma=\frac{1}{\rho}=\frac{e^2}{\Omega} \sum_\alpha \left
    (-\frac{\partial f}{\partial E_\alpha} \right )
    D_\alpha=e^2 N(E_F)\bar{D}.
\end{equation}
Thus, the quantum diffusivity is given explicitly by the
following expression
\begin{equation}\label{eq:statediff}
  D_\alpha^x=\pi \hbar \sum_{\alpha^\prime}
  |\langle \alpha|\hat{v}_x|\alpha^\prime \rangle|^2
\delta(E_\alpha-E_{\alpha^\prime}).
\end{equation}
In the semiclassical regime, $D_\alpha \rightarrow D_k=v_k
\ell_k/3$. However,  $D_\alpha$ can be used to characterize
transport even in the nonperturbative regime where semiclassical
concepts, like mean free path, loose their
meaning.~\cite{nikolic_qrho}
\begin{figure}
\centerline{\psfig{file=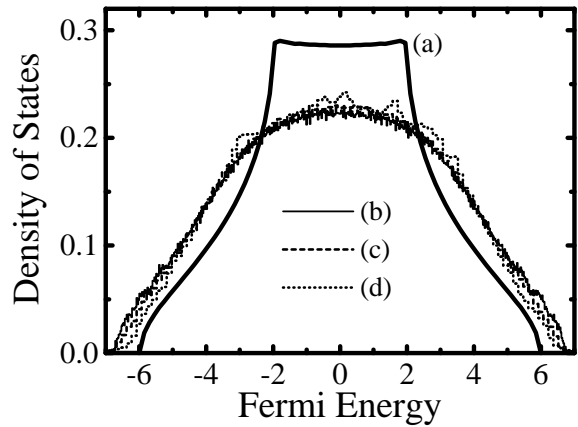,width=2.2in,angle=-90} }
\vspace{0.2in} \caption{The density of states of: (a) clean metal
($W=0$); (b) dirty metal ($W=6$ on a lattice $15 \times 15 \times
15$ averaged over 50 disorder configurations), obtained as $N(E)=2/{\Omega} \,
\sum_\alpha \delta(E-E_\alpha)$ from the exact eigenspectrum
$E_\alpha$ of a closed sample Hamiltonian; (c) dirty
metal [$W=6$ on a lattice $10 \times 10 \times 10$ averaged over
50 disorder configurations], obtained from the imaginary part
Eq.~(\ref{eq:dosim}) of the Green function~(\ref{eq:greenleads})
of an open sample; (d) is computed in the same way as (c) except that
smaller lattice $4 \times 4 \times 4$ is used.} \label{fig:clean_dos}
\end{figure}

To simplify calculations, I compute the quantum diffusivity
$\langle D_x (E_F) \rangle$ averaged over the disorder ($\langle
\ldots \rangle$ denotes averaging over an impurity ensemble), as
well as over a small energy interval. This is an additional
transport information, obviously related to
conductivity~(\ref{eq:kubo-diffusivity}), which is not usually
seen in the literature on disordered electron
physics.~\cite{prelovsek} It has been exploited in the transport
physics of glasses (e.g., to study the thermal conductivity in
amorphous silicon~\cite{allenglass}). The width $\eta$ of the
Lorentzian $\bar{\delta}(E_\alpha-E_{\alpha^\prime})$
in~(\ref{eq:statediff}) is chosen as some multiple of the local
average level spacing $\Delta(E_\alpha)$ in a small energy
interval around the eigenenergy $E_\alpha$. The method of
computing $\langle D_x (E_F) \rangle$ is as follows: a set of
eigenstates (the number of eigenstates is equal to the number of
lattice sites $N_s=N \times N_y \times N_z$) is obtained by
numerical diagonalization; $\mapsto$ for each eigenstate I compute
$D_\alpha^x$, where summation is going over all states
$|\alpha^\prime \rangle$ ``picked'' by the Lorentzian
$\bar{\delta}(E_\alpha-E_\alpha^\prime)$ (centered on $E_\alpha$)
in an energy interval of $3\eta$ around $E_\alpha$; $\mapsto$
finally, I average the diffusivities over the disorder and energy
interval defined by the bin of the size $\Delta E= 0.0225$.
\begin{figure}
\centerline{\psfig{file=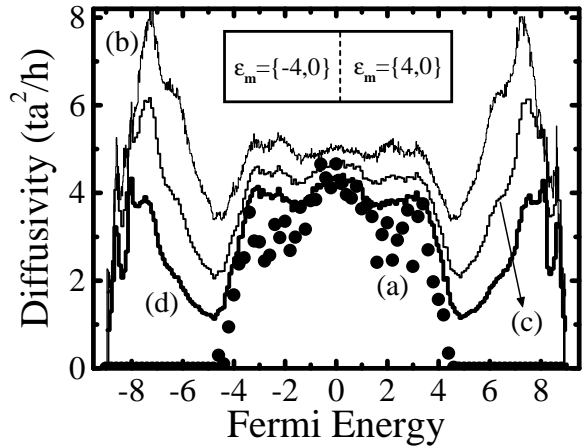,height=3.0in,angle=-90} }
\vspace{0.2in} \caption{Diffusivity $\langle D_x (E_F) \rangle$
for a metal junction composed of two disordered binary alloys
(modeled by TBH on a lattice $36 \times 8 \times 10$): (a)
computed using the Kubo formula~(\ref{eq:greenkubo}) in terms of
the Green function for the sample with attached leads; computed
from the Kubo formula in exact single particle eigenstate
representation~(\ref{eq:statediff}), where the width of the
Lorentzian broadened delta function is (b) $\eta=25 \Delta(E_F)$,
(c) $\eta=10 \Delta(E_F)$, and (d) $\eta=5 \Delta(E_F)$. Disorder
averaging is performed over 50 different samples.}
\label{fig:diffbainter}
\end{figure}
The efficient way of calculating quantum-mechanical average values
of some operator, like $\langle \alpha |\hat{v}_x|\alpha^\prime
\rangle$ appearing in the definition of eigenstate
diffusivity~(\ref{eq:statediff}), is to multiply three matrices
$\hat{\alpha}^\dag \cdot \hat{v}_x \cdot \hat{\alpha}$, where
$\hat{\alpha}$ is a matrix containing eigenvectors $|\alpha
\rangle$ as columns, and then take modulus squared of each matrix
element in such product. The number of operations in the
na\"{\i}ve calculation of the expectation values of operator,
where each of them is calculated separately, scales as $\sim
N_s^4$, while in the method presented above it scales as $\sim
N_s^3$ ($N_s \times N_s$ is the dimension of operator matrix).
This procedure becomes a natural choice once we understand that
it actually transforms the matrix of the operator $\hat{v}_x$
from the defining (site) representation into the representation
of eigenstates $|\alpha\rangle$. The end result of the
calculation, the average diffusivity $\langle D_x \rangle$, is
related to the conductivity through Einstein relation
\begin{equation}\label{eq:einstein}
 \sigma_{xx} =  e^2 N(E_F) \langle D_x(E_F) \rangle.
\end{equation}
Here $N(E_F)$ is the density of states (for both spin components)
evaluated at the Fermi level $E_F$.

I first calculate $\langle D_x (E_F) \rangle$  for the homogeneously
disordered sample (with binary disorder $\varepsilon_A=-2$,
$\varepsilon_B=2$) modeled on a lattice $18 \times 8 \times 10$,
as shown in Fig.~\ref{fig:diffbahomo}. To get an insight into the
microscopic features of the eigenstates used to evaluate the
KFESR (a few eigenstates around each $E_F$ determines the
transport properties at $E_F$), this figure also plots the inverse
participation ratio (IPR)~\cite{wegner}
\begin{equation}\label{eq:ipr}
I_2 = \Delta \left \langle \sum_{{\bf m},\alpha} \,
|\Psi_\alpha({\bf m})|^{4} \delta(E-E_{\alpha}) \right \rangle,
\end{equation}
averaged over disorder and energy $[ \Psi_\alpha({\bf m})=\langle
{\bf m} |\alpha \rangle ]$. This is the simplest single-number
measure of the degree of localization (i.e., the bigger the IPR
the more localized is the state is, e.g. IPR$=N_s$ corresponds to a
completely localized states on one lattice site). The IPR can also
be related to the average return probability~\cite{kramer} that
particle, initially launched in a state $|{\bf m} \rangle$
localized on a lattice site ${\bf m}$, will return to the same
site after a very long time (which is determined by its diffusive
properties encoded in $D_\alpha$). The second calculation, shown
in Fig.~\ref{fig:diffwhomo}, is for the homogeneous sample
described by the standard Anderson model where $\varepsilon_{{\bf
m}} \in [-W/2,W/2]$ is a uniform random variable. This results
are to be compared to the reference calculation based on the
mesoscopic Kubo formula.
\begin{figure}
\centerline{\psfig{file=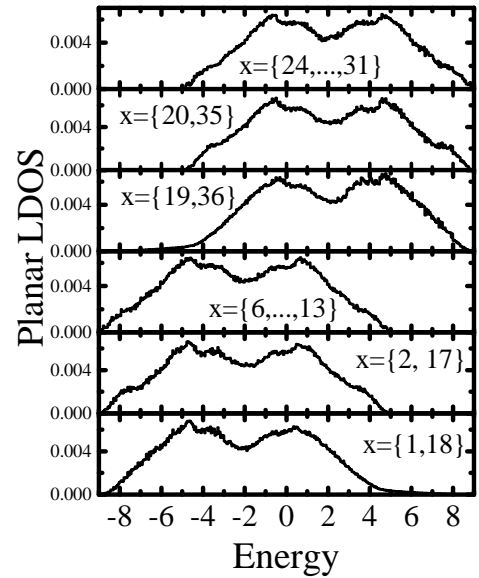,height=3.0in,angle=0} }
\vspace{0.2in} \caption{Local density of states (LDOS) integrated
over the $y$ and $z$ coordinates inside the metal junction
studied in Fig.~\ref{fig:diffbainter}. This ``planar LDOS''
$\rho(m_x,E)$ is computed from the exact eigenstates of the  TBH
using Eq.~(\ref{eq:ldosx}). The result is plotted after
$\rho(m_x,E)$ is averaged over several planes along the $x$-axis
(the planes used in this procedure are given on each panel).}
\label{fig:ldosx}
\end{figure}

Strictly speaking, the concept of eigenstate does not exist in
open systems. Such systems can be characterized by non-Hermitian
Hamiltonians,~\cite{datta} which do not conserve total
probability (cf. Sec.~\ref{sec:kubo}). This is a consequence of a
simple physical fact that an electron stays finite time within
the sample before escaping into the surrounding leads (i.e.,
initial discrete energy levels are broadened by the coupling to
the leads). Nonetheless, we can still get~\cite{janssen} the
density of states (DOS) from the imaginary part of the Green
function~(\ref{eq:greenleads})
\begin{equation}\label{eq:dosim}
  N(E_F)= \sum_{{\bf m}} -\frac{1}{\pi} {\rm Im} \,
  \hat{G}^r({\bf m},{\bf m};E_F).
\end{equation}
That such DOS of an open 3D system is indistinguishable from the
one computed from the distribution of energy eigenvalues,
$N(E_F)=(2/\Omega) \, \langle \sum_\alpha \delta(E_F-E_\alpha)
\rangle$, is shown in Fig.~\ref{fig:clean_dos}. This is in spite
of the fact that leads strongly perturb the edges of the system,
which is clearly exhibited  only on the smallest lattice in
Fig.~\ref{fig:clean_dos}. Thus, the average diffusivity $\langle
D_x (E_F) \rangle$ can be computed in a straightforward manner
from the Einstein relation~(\ref{eq:einstein}) where conductivity
is formally expressed from the disorder-averaged conductance
$\sigma_{xx}=\langle G_{xx}\rangle L_x/S$ given by the mesoscopic
Kubo formula.
\begin{figure}
\centerline{\psfig{file=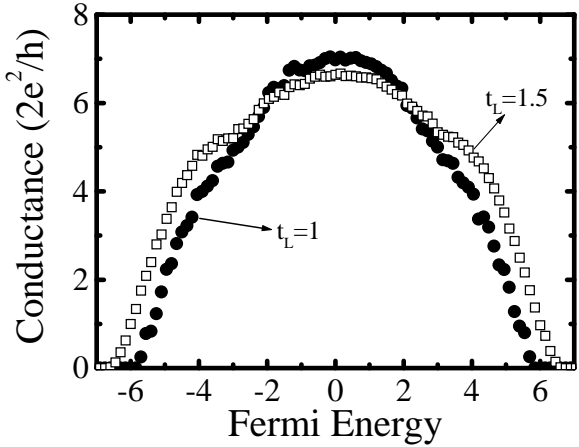,height=3.0in,angle=-90} }
\vspace{0.2in} \caption{Two-probe conductance of a disordered
conductor (averaged over 50 samples) modeled by the Anderson
model with diagonal disorder $W=6$ on a simple cubic lattice
$10^3$. The computation is done using the mesoscopic Kubo
formula~(\ref{eq:greenkubo}) for the finite sample attached to
two semi-infinite leads characterized by two different values of
hopping integral $t_{\rm L}$. Note that the conductance vanishes
at $|E|=6t$ (band edge in a clean sample) when $t_{\rm L}=t$.}
\label{fig:homolead}
\end{figure}

In both calculations for the homogeneous samples it appears that
discrepancy between the Kubo formula in single particle
representation~(\ref{eq:kuboesr}) and the exact method, based on
the formula~(\ref{eq:greenkubo}) for the {\em sample + lead}
system, is only numerical. In fact, the difference is very small
in the disordered binary alloy and a bit larger in the Anderson
model with continuous disorder. It originates from the ambiguity
in using the width $\eta$ of the broadened delta function. Namely,
nonzero $\eta$ effectively introduces inelastic scattering as an
uncorrelated random event.\cite{kirkpatrick} The increase of the
diffusivity close to the band edges of diagonally disordered
Anderson model (cf. Fig.~\ref{fig:diffwhomo}) was seen long time
ago in the direct simulations of the wave function diffusion
performed in early days of localization theory.~\cite{prelovsek}
\begin{figure}
\centerline{\psfig{file=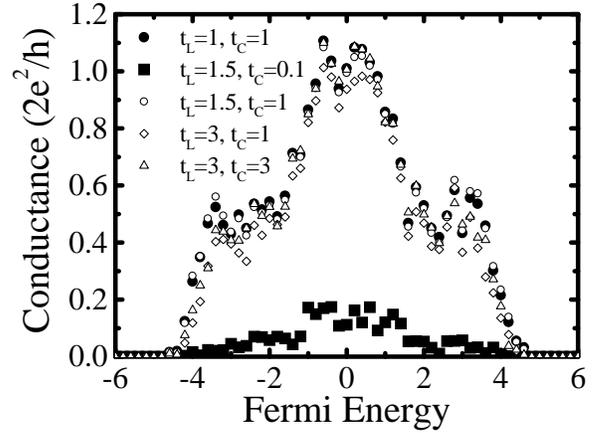,height=3.0in,angle=-90} }
\vspace{0.2in} \caption{Two-probe conductance (averaged over 50
samples) of the same dirty metallic junction sample as in
Fig.~\ref{fig:diffbainter}, but attached to two semi-infinite
ideal leads with several different hopping integral $t_{\rm L}$ or
lead-sample couplings $t_{\rm C}$ (hopping integral in the sample
sets the unit of energy $t=1$). The computation is based on the
mesoscopic Kubo formula Eq.~(\ref{eq:greenkubo}).}
\label{fig:junctionlead}
\end{figure}

The same analysis is repeated for a junction (introduced at the
beginning of this section) which is composed of two different
disordered binary alloys on each side. The result is shown in
Fig.~\ref{fig:diffbainter}. Large fluctuations of the diffusivity
(i.e., quantum conductance $G$ from which diffusivity is computed
at specific $E_F$) are caused by the conductance being of the
order of $2e^2/h$ (Fig.~\ref{fig:junctionlead}). Such
fluctuations are less obvious in the result based on the KFESR
because of the extra averaging over energy provided by the
Lorentzian broadened delta function. Here the discrepancy between
the two different methods is not only quantitative, but the
KFESR~(\ref{eq:kuboesr}) shows nonzero diffusivity (and thereby
conductivity, since global DOS is nonzero all throughout the band)
at Fermi energies at which there are no states on one side of the
junction which can carry the current.~\cite{andersonhub} The
result persist even when the width $\eta$ of the Lorentzian
broadened delta function is decreased. Therefore, it is not an
artifact of the numerical trick used to evaluate KFESR. The
states which have nonzero amplitude all throughout the junction
cease to exist at $|E_b| \sim 4.7$, which is seen by inspecting
the local density of states (LDOS), integrated over $y$ and $z$
coordinates
\begin{eqnarray}\label{eq:ldosx}
  \rho(m_x,E) & = & \sum_{m_y,m_z} \rho({\bf m},E) = \nonumber \\
  &=& \sum_{m_y,m_z} \! \sum_\alpha \, |\Psi_\alpha({\bf m})|^2 \, {\delta}
(E-E_\alpha).
\end{eqnarray}
Such ``planar LDOS'' along the $x$-axis is plotted in Fig.~\ref{fig:ldosx}.
It changes abruptly while going from one
side of the junction to the other side (except for the small
tails near the interface). On the other hand, Fig.~\ref{fig:diffbainter}
explicitly demonstrates that Kubo
formula~(\ref{eq:greenkubo}) for an open finite-size sample
plugged between ideal semi-infinite leads correctly describes
this junction. Namely, the diffusivity vanishes at the same point
at which LDOS goes to zero. It should be emphasized that once the
leads are attached, two new interfaces (lead-sample) in the
problem arise. The Landauer-B\" uttiker scattering approach to
transport intrinsically takes care of these boundaries by
considering the realistic finite-size system where electrons can
leave or enter through some surfaces into the rest of the circuit
represented by the leads.~\cite{landimry,andersonbook} Thus,
mesoscopic developments have clarified the way of proper
application of Kubo formulism to finite-size samples (which is
ultimately related to the eternal puzzle of the origins of
dissipation in conservative systems, technically the only ones
appearing in the Kubo analysis).~\cite{landphil} This in turn has justified
heuristic arguments of Landauer on a rigorous basis.~\cite{fisher,baranger-kubo}

The conductance of an infinite ({\it sample + leads}) system will
go to zero at the band edge of a clean lead $|E_b|=6t$ if we use
the same hopping parameter in the lead $t_{\rm L} = t$ as in the
disordered sample. This stems from the fact that are no states in
the leads which can carry the current for Fermi energies $|E_F| >
6t$ outside of the clean TBH band. Technically, the self energy
$\hat{\Sigma}^r$ is real at these energies, which leads to ${\rm
Im}\, \hat{G}$ in~(\ref{eq:greenkubo}) being zero. Thus, the
conductance of the whole band of disordered sample cannot be
computed unless we increase $t_{\rm L}$ in the leads. This is
illustrated in Fig.~\ref{fig:homolead} for the homogeneous sample
described by the Anderson model where band edge of the disordered
sample $|E_b| > 6t$ lies outside of the clean metal band defined
by crystal symmetry. Thus, a natural question arises when using
$t_{\rm L},\, t_{\rm C} \neq t$: how sensitive is the conductance
to the properties of leads or the sample-lead coupling? This
problem resembles the quantum measurement problem because
semi-infinite leads can also be viewed as a macroscopic apparatus
necessary for measurement.~\cite{efetovbook,nikolic_qpc} This is
further stressed in multi-probe geometries~\cite{datta} where
extra leads are introduced to measure the voltages along the
sample (besides the two leads, used here, through which the
current is fed). The measured conductance is then property of
both the sample and the measurement geometry, which is one of the
reasons for mesoscopic physics to employ only sample- and measurement
geometry-specific quantities, like quantum conductance, instead
of intensive quantity like conductivity.~\cite{stonebook} From a
transport point of view, it is clear that lead-sample interfaces
introduce extra scattering for different effective masses or Fermi velocities
in the sample and the lead. Thus, the {\em exactness} of the
conductance calculated here is, in fact, a feature of the whole {\em sample + leads}
setup in Fig.~\ref{fig:setup} (that is akin to any kind of measurement
in quantum mechanics), and it is important to confirm
that general conclusions about the relationship between
different Kubo formulas do not depend on particular values
of the chose parameters for the leads.

It is understood~\cite{hans} that if broadening of the energy
levels due to the coupling to semi-infinite leads is greater than
the Thouless energy $E_{\rm Th} \simeq \hbar {\mathcal D}/L^2$,
(${\mathcal D}=v_F \ell/3$ is the classical diffusion constant),
then level discreteness of an isolated sample is unimportant and
$G$ will be independent of the properties of leads. This limit
corresponds to an ``intrinsic'' (Thouless) conductance
$G/G_Q=2\pi E_{\rm Th}/\Delta$ ($\Delta$ is the mean level
spacing) that is smaller than the conductance determined by the
lead-sample contact.~\cite{mackinnon} It requires disordered
enough sample~\cite{nikolic_qpc} and leads of the same cross
section as that of the sample.~\cite{mackinnon} Thus, even though
attaching the sample to the leads brings dramatic changes
(discrete spectrum is changed into continuous and new boundary
conditions are introduced), the conductance is determined by the
sample properties only, and dissipation in the reservoirs does not
enter into computational algorithm. This dependence is studied in
Fig.~\ref{fig:junctionlead} by looking at the conductance of our
model junction as a function of the hopping in the leads $t_{\rm
L}$ and the coupling $t_{\rm C}$. The conductance is virtually
independent of $t_{\rm L}$, which is a consequence of the
smallness of the disordered junction conductance (as discussed
above). It goes down drastically with decreasing of the coupling
$t_{\rm C}$ (the same behavior is anticipated when $t_{\rm L}$ is
increased substantially because of the increased reflection at
the lead-sample interface).

\section{Transport through strongly disordered interface}
\label{sec:interface}

In this section we analyze quantum transport properties of a
single dirty interface whose dimensionless conductance $G/G_Q$ is
much smaller than the number of conducting channels $N_{\rm ch}$
(i.e., much smaller than  the conductance of a ballistic
conductor of the same cross section). For practical purposes,
interface can be defined as a any scattering region whose
thickness is sufficiently shorter than~\cite{schep-bauer}
$\lambda_F$ (in bulk conductors $L \gg \lambda_F$). Here we look
at a geometrical plane of atoms as a model of interface in the
strict sense. Furthermore, the evolution of transport from single
dirty interface to strongly disordered thin slab (i.e., ``thin
Anderson insulators'', since stacking together enough finite-size
interfaces, with the disorder strength chose here, would lead to
a bulk Anderson insulator having exponentially small
conductance). Such thin slabs are a more likely element in
experimental circuits.~\cite{yehuda} These problems are not only
conceptual, namely to understand the difference between the
transport in bulk conductors and interfaces, but it has been
brought about by the understanding of crucial effects the
interface scattering can have in the CPP transport
experiments~\cite{cppgmr} on GMR magnetic multilayers (in GMR the
added complication is spin-dependent interface
resistance~\cite{stiles} which dominates the resistance and
magnetoresistance for not too large layer
thicknesses~\cite{qyang}).

The importance of interface scattering in many areas of metal and
semiconductor physics has been realized in a plethora of research
papers since the seminal work of Fuchs.~\cite{fuchs} They are
mainly concerned with the transport parallel to impenetrable
rough interface, while recent have also risen interest in the
transport normal to the interface (CPP geometry). Because the
nature of the transport relaxation time in inhomogeneous systems
is not well understood,~\cite{gijs} the first step is to
understand properties of a single interface before studying them
as a part of some more complicated circuit like those in
Sec.~\ref{sec:multilayer}. For example, the properties of a
single interface cannot be described in terms of the Boltzmann
conductivity $\sigma_B=ne^2\tau/m$, i.e.,~using the elastic mean
free path $\ell=v_F \tau$ (or transport mean free time $\tau$)
familiar from the bulk metallic conductors whose conductivity is
dominated by the semiclassical effects.
\begin{figure}
\centerline{\psfig{file=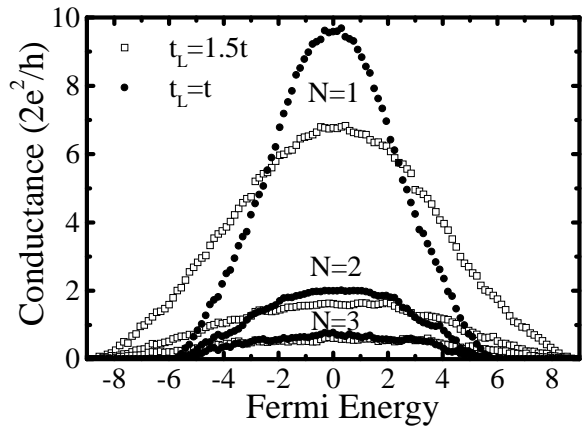,width=2.2in,angle=-90} }
\vspace{0.2in} \caption{Conductance of a single dirty interface
($N=1$) and thin slabs composed of two ($N=2$) or three ($N=3$)
such interfaces (modeled by the Anderson model with diagonal
disorder $W_I=30$ on a lattice $N \times 10 \times 10$). The
calculation is performed using different values of the hopping
parameter $t_{\rm L}$ in the attached leads, and the results are
averaged over $200$ realizations of disorder.}
\label{fig:interfaceg}
\end{figure}
\begin{figure}
\centerline{\psfig{file=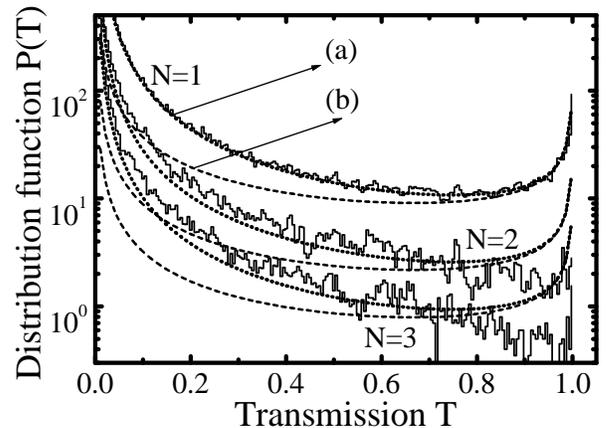,width=2.2in,angle=-90} }
\vspace{0.2in} \caption{Distribution function $P(T)$ of the
transmission eigenvalues (at half-filling $E_F=0$) for a single
disordered interface $N=1$, and thin slabs composed of two $N=2$
or three $N=3$ such interfaces, modeled by the Anderson model
with $W_I=30$ on a cubic lattice $N \times N_y \times N_z$. The
disorder-average is taken over an ensemble of $1000$ conductors.
The analytical functions plotted are: (a) $P(T)=(G/2G_Q) \,
1/(T^{3/2}\sqrt{1-T})$, and (b) Dorokhov distribution
$P(T)=(G/2G_Q) \, 1/(T\sqrt{1-T})$.} \label{fig:interfacett}
\end{figure}

Lacking enough experimental information, the simple theoretical
models for the interface effects on electron propagation assume
diffuse scattering at interdiffused atoms or interfacial
roughness~\cite{schep2} (these free electron theories omit the
complex electronic band structure of transition metals appearing
in realistic GMR samples). Even disorder-free interface can have a
nonzero resistance, because of mismatch of the crystal potential
and band structures.~\cite{gijs} Here we are interested only in
generic properties of mesoscopic transport through interfaces
that do not depend on material-specific
details.~\cite{schep-bauer} Therefore, the interface roughness is
modeled here by the short range random scattering
potential~\cite{bauermetal} generated by the impurities located
on the sites of a square lattice $1 \times N_y \times N_z$ and
with strong disorder $W_I=30$ in the corresponding TBH
Eq.~(\ref{eq:tbh}). The bulk conductor composed of such
interfaces (stacked in parallel and coupled with nearest neighbor
hopping $t$) is an Anderson insulator, because all states are
localized already for~\cite{slevin} $W_c \simeq 16.5$. To compute
the CPP transport properties  using the mesoscopic Kubo
formula~(\ref{eq:greenkubo}), the interface is placed in our
standard computational setup between the two semi-infinite
disorder-free leads. Thus, the conductance is computed for an
atomic monolayer of a disordered material inside an infinite
clean sample of a finite cross section shown in
Fig.~\ref{fig:setup}. Also computed is the conductance of a thin
slab composed of two (i.e., 3D conductor modeled on the lattice $2
\times N_y \times N_z$) and three sheets ($3 \times N_y \times
N_z$) of the same disordered material, as shown in
Fig.~\ref{fig:interfaceg}. The microscopic origin of different
terms contributing to the interface resistance was traced back to
both specular and diffusive scattering in the plane (where
diffusive scattering can even open additional channels for
electron transport, thereby increasing the
conductance).~\cite{gijs}

Following the discussion in Sec.~\ref{sec:metaljunc}, the
influence of the leads on the interface conductance is checked by
using two different hopping integrals $t_{\rm L}$ (compare to
Fig.~\ref{fig:homolead}). Here the analysis based on comparison
of relevant energy scales does not work (i.e., one can not use
bulk material concepts, like $E_{\rm Th}$). Plausibly, I find that
leads affect the conductance of the interface much more than the
conductance of a bulk disordered conductor (with similar value of
disorder-averaged conductance).
\begin{figure}
\centerline{\psfig{file=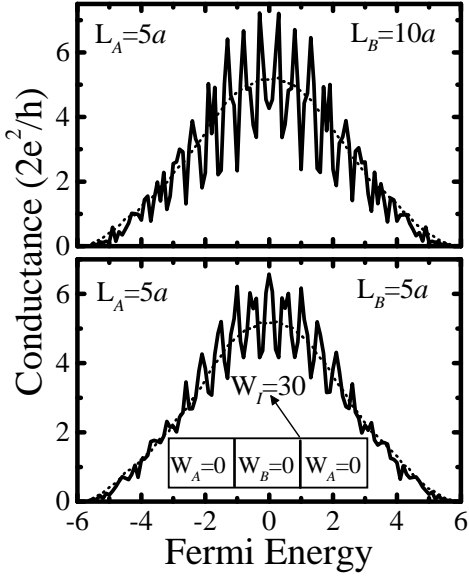,height=3.0in,angle=0} }
\vspace{0.2in} \caption{The disorder-averaged (over 200
configurations) conductance of a mesoscopic multilayer, composed
of dirty interfaces and clean bulk conductors, modeled on a
lattice $17 \times 10 \times 10$. The results are obtained from:
the mesoscopic Kubo formula~(\ref{eq:greenkubo}) applied to the
whole multilayer (solid line); and using the semiclassical
resistor model, $G_{\rm RM}=(2R_I-R_{\rm QPC})^{-1}$ (dotted
line), where individual resistances are computed also from
Eq.~(\ref{eq:greenkubo}) and summed accordingly (the meaning of
$R_{\rm QPC}$ is explained in Fig.~\ref{fig:layerelem}).}
\label{fig:mlbal2box}
\end{figure}

Mesoscopic transport methods give the possibility not only to
compute the conductance, but also to use the picture of
conducting channels and their quantum-mechanical transmission
properties. From a pragmatical point of view, one does not need
these fully quantum techniques to study the transport in
macroscopic conductors which are usually dominated by
semiclassical physics. Nevertheless, the study of the
distribution of transmission probabilities, which requires
phase-coherent transport, enhances our insight into the
conduction processes in electronic systems. The diagonalization
of a Hermitian matrix ${\bf t} {\bf t}^\dag$, which defines the
conductance through the Landauer formula,
\begin{mathletters}\label{eq:ttlandauer}
\begin{eqnarray}
  G & = & \frac{2e^2}{h} \, \text{Tr} \, ({\bf t t}^{\dag}) =
  \frac{2e^2}{h} \sum_{n=1}^{N_y N_z} T_n, \\
  {\bf t} & = & 2 \sqrt{-\text{Im} \, \hat{\Sigma}_L} \, \hat{G}^{r}_{1 N}
  \sqrt{-\text{Im}\, \hat{\Sigma}_R},
  \label{eq:t}
\end{eqnarray}
\end{mathletters}
introduces a set of transmission eigenvalues $0 \leq T_n \leq 1$
\begin{figure}
\centerline{\psfig{file=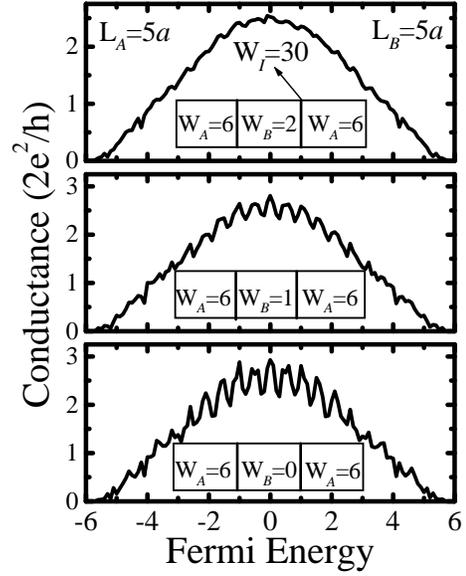,height=3.0in,angle=0} }
\vspace{0.2in} \caption{The disorder-averaged (over 200
configurations) conductance of a mesoscopic multilayer, composed
of strongly disordered interfaces and (quasi)ballistic bulk
conductors (e.g., $\ell \approx 9$ at half-filling for $W=2$),
modeled on a lattice $17 \times 10 \times 10$. The results are
obtained from the mesoscopic Kubo formula~(\ref{eq:greenkubo})
applied to the whole multilayer as a single quantum-coherent
conductor.} \label{fig:mlbal3box}
\end{figure}
for each realization of disorder. The incident flux concentrated
in channel $| p \rangle$ will give the wave function in the
opposite lead $\sum_q {\bf t}_{pq} |q \rangle$, where ${\bf t}$
is the transmission matrix. Here $\hat{G}^{r}_{1 N}$  is the
matrix block, which connects  layers $1$ and $N$ of the sample,
of the full retarded real-space Green function
matrix~(\ref{eq:greenleads}). The distribution function of $T_n$
is defined as
\begin{equation}\label{eq:pt}
  P(T)=\left \langle \sum_n \delta (T-T_n) \right \rangle.
\end{equation}
Using $P(T)$, the disorder-average value of any quantity that can
be written down in the form of linear statistics $a(T)$ can be
expressed as~\cite{carlo_rmt}
\begin{equation}\label{eq:linearstat}
  \langle A \rangle = \left \langle \sum_{n=1}^{N_{\rm ch}} a(T_n)
  \right \rangle = \int\limits_0^1 d T \, a(T) P(T).
\end{equation}
For example, the conductance is described by a linear statistics
$g(T)=T$. Contrary to the na\"{\i}ve expectations, stemming from
comparison with Drude-Boltzmann conductance $G \simeq G_Q N_{\rm
ch}\ell/L$, that transmission of every channel is $T \simeq
\ell/L$ in the metallic diffusive ($G_Q N_{\rm ch} \gg G \gg
G_Q$) bulk conductor, it was shown by Dorokhov~\cite{dorokhov}
that in a homogeneous multichannel wire geometry
\begin{equation}\label{eq:dorokhov}
  P_{\rm D}(T)=\frac{G}{2G_Q} \, \frac{1}{T \sqrt{1-T}},
  \\\ \cosh^{-2} \left( \frac{N_{\rm ch}}{g}
  \right) < T < 1.
\end{equation}
The cutoff~\cite{schep-bauer} at small $T$ is such that $\int_0^1
d T \, P(T) = N_{\rm ch}$ ensuring that averages of the first and
higher order moments of $T$ are not affected (on the proviso
$N_{\rm ch} \gg G/G_Q$). The Dorokhov distribution is
universal---it depends only on the disorder-averaged conductance
$G$ and not on the details of disorder, dimension, shape of the
sample, carrier density, spatial resistivity distribution and
other sample-specific properties.~\cite{carlo_rmt} Thus, $P(T)$
is bimodal distribution function meaning that: most of $T_n$ are
either $T_n \simeq 0$ (``closed'' channels) or $T_n \simeq 1$
(``open'' channels). This has important consequences when
calculating linear statistics other than the conductance since we
can get conductance (first moment of the distribution) without
really knowing the details of $P(T)$ (e.g., the higher moments are
probed in the case~\cite{carlo_rmt} of shot-noise power, Andreev
conductance of metal/supercondcutor interface, etc.).
\begin{figure}
\centerline{\psfig{file=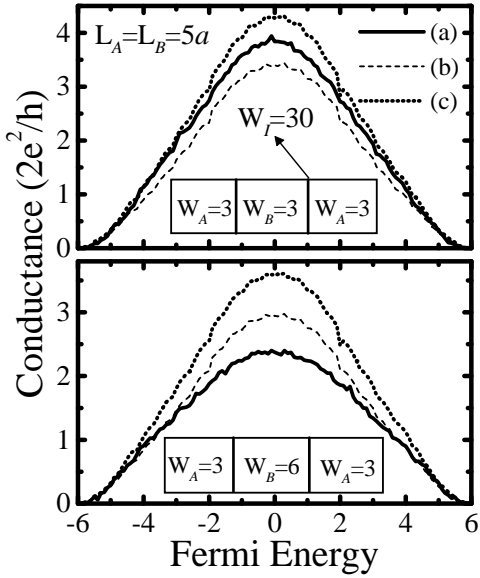,height=3.0in,angle=0} }
\vspace{0.2in} \caption{The disorder-averaged (over 200
configurations) conductance of a mesoscopic multilayer, composed
of strongly disordered interfaces and weakly disordered bulk
conductors, modeled on a lattice $17 \times 10 \times 10$ ($\ell
\approx 4a$ at half-filling for $W=3$). The results are obtained
from: (a) mesoscopic Kubo formula~(\ref{eq:greenkubo}) evaluated
for the whole multilayer; following the resistor model
Eq.~(\ref{eq:resistor}), (b) $G_{\rm
RM}^\prime=(2R_A+R_B+2R_I)^{-1}$, and (c) $G_{\rm
RM}=(2R_A+R_B+2R_I-4R_{\rm QPC})^{-1}$.} \label{fig:layerdifw3}
\end{figure}

The simple counting of the number of $T_n$ in each bin along the
interval $[0,1]$ gives allows us to obtain the distribution
function $P(T)$ (in this procedure the delta function
in~(\ref{eq:pt}) is effectively broadened into a box function
$\bar{\delta}(x)$ equal to one inside each bin).
Figure~\ref{fig:interfacett} plots $P(T)$ for the dirty interface
and the two thin slabs. The result is compared to $P_{\rm D}(T)$
of Eq.~(\ref{eq:dorokhov}) and the one which describes the
numerical data
\begin{equation}\label{eq:rhotmy}
  P_{\rm fit}(T) = \frac{G}{2 G_Q} \, \frac{1}{T^{3/2} \sqrt{1-T}}.
\end{equation}
This formula is, up to a numerical factor, the same as the
analytical prediction of Schep and Bauer~\cite{schep-bauer} for a
single dirty interface
\begin{equation}\label{eq:schep-bauer}
  P_{\rm SB}(T)=\frac{G}{\pi G_Q} \, \frac{1}{T^{3/2} \sqrt{1-T}},
\end{equation}
Thus, interfaces  belong to a universality class different from
that of diffusive  bulk conductors characterized by the Dorokhov
distribution $P_{\rm D}(T)$. However, it seems that distribution
is valid even for thin slabs whose thickness $L$ is grater than
$\lambda_F$, but is smaller than the localization radius (which
in our case can be estimated in the pedestrian way as the
thickness at which conductance vanishes). This is in compliance
with experimental confirmation of $P_{\rm SB}(T)$ in a uniform
Nb/AlO$_x$/Nb Josephson junction, whose subharmonic gap structure
in the $I-V$ characteristic is extremely sensitive to the number
of conducting channels at $E_F$ and their transmission
probabilities~\cite{bardas}, since the thickness of the realistic
AlO$_x$ barrier is bigger than $\lambda_F$ in the junction
electrodes (Ref.~\onlinecite{yehuda} gives a simple derivation of
the Schep-Bauer distribution without using the assumption $L \ll
\lambda_F$). In fact, even our extremely high disorder input in
the standard Anderson model generates surface conductance in the
band center, $g_I \simeq 9 \cdot 10^{10} \ \Omega^{-1}$cm$^{-2}$
(for $a=3$ \AA), that is much higher than $10^8
\,\Omega^{-1}$cm$^{-2}$ reported for the measured value in
Ref.~\onlinecite{yehuda} (suggesting that thin slab is more
likely to play to role of a barrier in these Josephson junction
than the strict geometrical plane). While being intriguing concept
in disorder electron physics, universality can be frustrating for
the device engineers. Not all features of the transport through
dirty interface are universal.~\cite{schep-bauer}
\begin{figure}
\centerline{\psfig{file=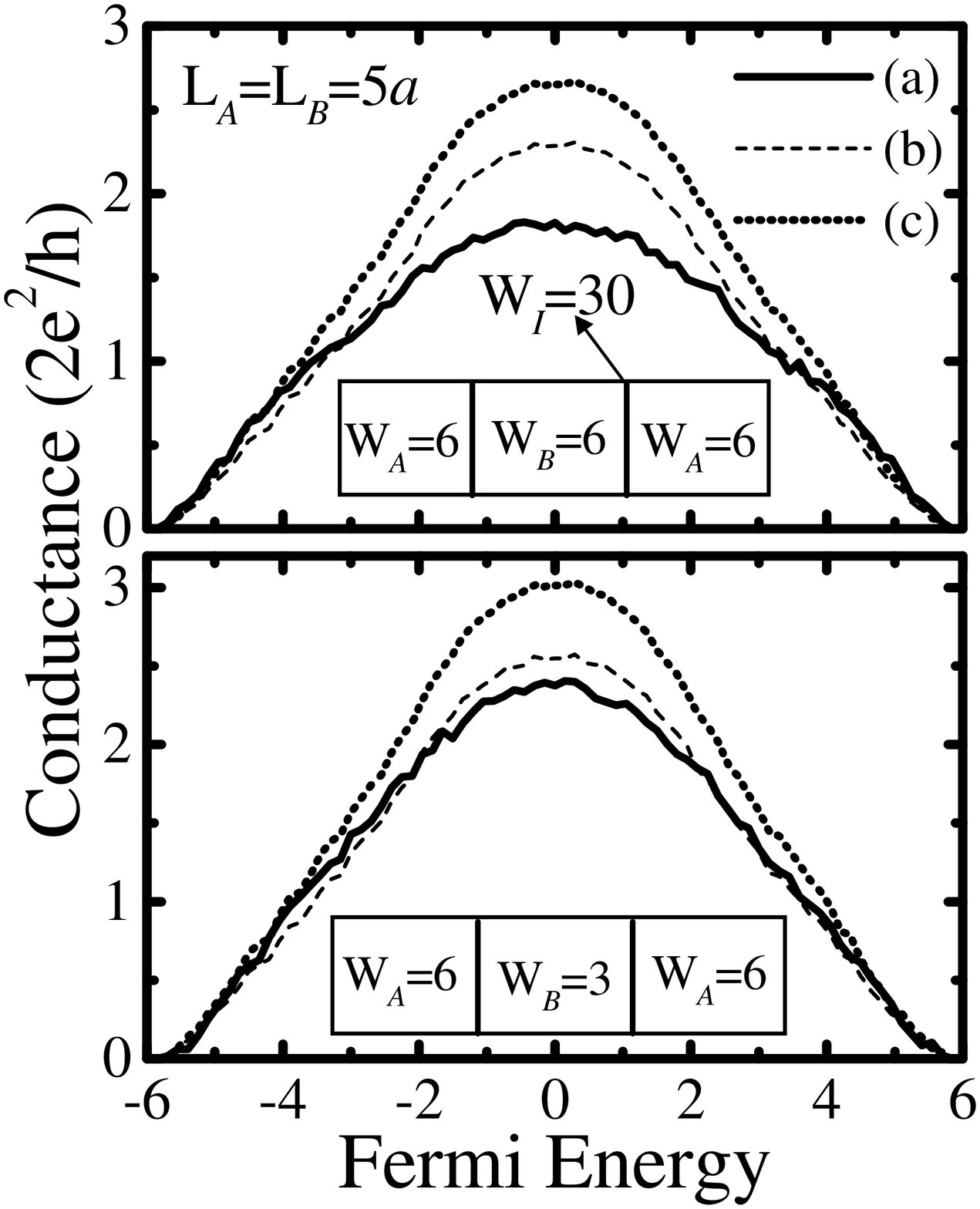,height=3.0in,angle=0} }
\vspace{0.2in} \caption{The disorder-averaged (over 200
configurations) conductance of a mesoscopic multilayer, composed
of strongly disordered interfaces and weakly disordered bulk
conductors, modeled on a lattice $17 \times 10 \times 10$. The
results are obtained from: (a) mesoscopic Kubo
formula~(\ref{eq:greenkubo}) evaluated for the whole multilayer;
following the resistor model  Eq.~(\ref{eq:resistor}), (b)
$G_{\rm RM}^\prime=(2R_A+R_B+2R_I)^{-1}$, and (c) $G_{\rm
RM}=(2R_A+R_B+2R_I-4R_{\rm QPC})^{-1}$.} \label{fig:layerdif}
\end{figure}

\section{Transport through multilayers} \label{sec:multilayer}

Armed with the knowledge of transport properties of dirty
interfaces and metallic homogeneous layers, we can now undertake
the study of circuits composed of such elements. The main feature
of our circuits is that they are of nanoscale size ($\sim$ few
$\lambda_F$) and fully phase-coherent (i.e., effectively at zero
temperature). The choice of disorder and size of the system is
driven by the interest here to explore the patterns of breaking
(under the influence of quantum effects) of a simple description
of the multilayer in terms of classical resistor network.
Therefore, I study such deviations from semiclassics by computing
the exact zero-temperature conductance of several multilayers, as
well as of their components, using mesoscopic Kubo
formula~(\ref{eq:greenkubo}). This computation takes into account
all quantum interference and quantum size effects from the onset.
In general, resistors can be added according to the classical
Ohm's law only when their size is larger than the dephasing length
$L_\phi$ (in this case the quantum features of diffusion can still
enter through the transport properties of individual
phase-coherent units of size $L_\phi$, example being the weak
localization effect~\cite{wl} at finite temperatures). For
example, at high enough temperatures the system can be partitioned
into the cubes of a macroscopic size $L_\phi$ where quantum
interference between wave functions scattered in different cubes
can be neglected---this makes it possible then to define an
intensive quantity,~\cite{janssen} like conductivity, by solving a
classical random resistor network problems.~\cite{kaveh} The
quantum composition law for quantum resistors in a chain is more
complicated since it contains a phase variable depending on the
characteristics of all scatterers.~\cite{gang4_seq,buttiker86}
Thus, because of quantum interference effects (like conductance
fluctuations~\cite{ucf}) phase-coherent resistors do not add in
series, and even for homogeneous samples resistance as a function
of length is not a self-averaging quantity (therefore requiring
disorder-averaging to restore the Ohmic
scaling~\cite{nikolic_qrho,todorov}). In the Landauer-B\" ttiker
scattering formalism, resistor network description corresponds to
a semiclassical concatenation of the scattering matrices of
individual units (${\bf t}$ in~(\ref{eq:ttlandauer}) is just one
block of a full scattering matrix~\cite{carlo_rmt}), i.e., the
concatenation of ``probability scattering matrices'' of
successive disordered regions (obtained by replacing each element
of the scattering matrix by its squared modulus~\cite{cahay}).

The multilayers studied here are composed of three bulk conductors
joined through two dirty interfaces. The whole structure is
modeled by the TBH Eq.~(\ref{eq:tbh}) on a lattice $17 \times 10
\times 10$ where sixth and twelfth monoatomic layers contain the
same interface studied in Sec.~\ref{sec:interface}. The disorder
strength within the plane of the interface is fixed to $W_I=30$,
while disorder inside the bulk layers (composed of five
monolayers) is varied. The disorder strength is taken to be the
same in two outer layers where diffusive bulk scattering takes
place. This type of multilayer can be viewed as a period of an
infinite $A/B$ multilayer:~\cite{schep2} layer of material $A$ on
the outside (of resistivity $\rho_A$ and total thickness
$2L_A=10a$, where $a$ is the lattice spacing) and interlayer
material $B$ between the interfaces (of resistivity $\rho_B$ and
thickness $L_B=5a$). For example, for the chosen disorder
strengths $W=3$ and $W=6$  the corresponding bulk material
resistivities at half-filling are $\rho \simeq 130\, \mu
\Omega$cm and $500\, \mu \Omega$cm (assuming $a=3 {\text \AA}$),
respectively. I neglect any potential step at the interface
(caused by the conduction band shift at the
interface~\cite{gijs}). Such multilayers are often described
semiclassicaly in terms of the resistor model~\cite{levy}
\begin{equation}\label{eq:resistor}
  R_{\rm RM}= N_b \left (\rho_A \frac{2L_A}{S} + \rho_B \frac{L_B}{S} + 2 R_I \right),
\end{equation}
where $R_{\rm RM}$ is the total multilayer resistance, $N_b$ is
the number of bilayers (I study below just one multilayer period
$N_b=1$), and $R_I$ is the interface resistance (evaluated in
Sec.~\ref{sec:interface} in Fig.~\ref{fig:interfaceg}). Thus, the
resistor model treats both bulk and interface resistances as
semiclassical elements of a circuit in which resistors add
Ohmically in a series. Nevertheless, quantum diffusion can be
important inside each individual resistor element, as discussed
above.~\cite{datta} From the measurement of $R_{\rm RM}$ as a
function of the layer thickness, the bulk and interface
resistances can be extracted experimentally. When quantum
interference effects become important in the CPP transport, this
picture breaks down.
\begin{figure}
\centerline{\psfig{file=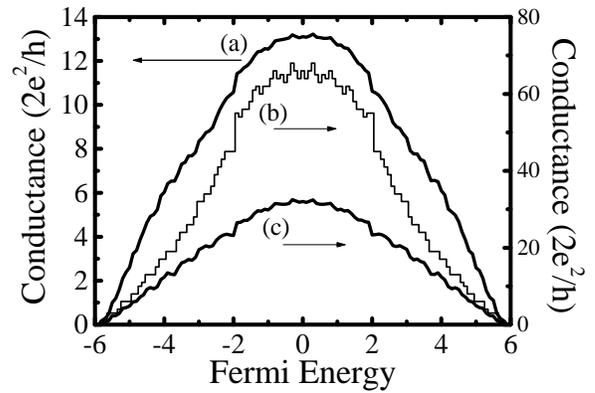,height=3.0in,angle=-90} }
\vspace{0.2in} \caption{Conductance of individual resistors
comprising the multilayers of Sec.~\ref{sec:multilayer}:
disordered conductors modeled by the Anderson model a lattice $5
\times 10 \times 10$ with diagonal disorder strength, (a) $W=6$,
and (b) $W=3$, and quantum point contact conductance ($1/R_{\rm
QPC}$) of a clean sample modeled on the lattice with the same
cross section (i.e., on  a lattice supporting $100$ conducting
channels, where the maximum of open channels with $T_n=1$ at the
Fermi level in such 3D ballistic conductors~\cite{nikolic_qpc} is
$68$ at $|E_F| \simeq 0.3$).} \label{fig:layerelem}
\end{figure}

The quantum (i.e., zero-temperature) conductance of the
multilayers is computed using a standard tool throughout the
paper---mesoscopic Kubo formula of Eq.~(\ref{eq:greenkubo}) where
whole multilayer is treated as a single phase-coherent unit. This
approach intrinsically takes into account finite-size effects in
the problem,~\cite{landimry} as emphasized in
Sec.~\ref{sec:kubo}, as well as all single-particle quantum
interference effects. In all calculations the hopping integrals
throughout the disordered sample and in the leads are the same
($t_{\rm L}=t_{\rm C}=t$). This means that no additional
scattering, discussed in Sec.~\ref{sec:metaljunc}), is generated
at the lead-sample interface.~\cite{nikolic_qpc} The remaining
resistance can sometimes be interpreted as a series addition of
two quantum resistors. Namely, the semiclassical limit of the
two-probe Landauer or Kubo formula for conductance, obtained
e.g., from the stationary-phase approximation~\cite{semilandauer}
of the Green function expression~(\ref{eq:ttlandauer}) for the
transmission of the sample, leads to
\begin{equation}\label{eq:semilandauer}
  \langle G \rangle^{-1} =R_{\rm QPC}+ \rho \frac{L}{S},
\end{equation}
which should be valid for not too strong scattering on
impurities. The ``contact'' resistance~\cite{sharvin} $R_{\rm
QPC}=\pi \hbar/e^2 N_{\rm ch}$ is nonzero even for ballistic
conductor because a finite cross section can carry only finite
currents (the voltage drop in this case occurs at the
lead-reservoir interface). Similar interpretation was given for
the interface resistance,~\cite{schep2} and is expected to be
valid also for the interfaces embedded into a multilayer if there
is no coherent scattering between adjacent interfaces (e.g., when
such effects are destroyed by sufficiently strong scattering in
the bulk). These formulas naturally lead to the resistor model
where different interfaces and bulk layers contribute to the CPP
resistance as resistors in series. It is often assumed that
$R_{\rm QPC}$ can be neglected when compared to usually much
higher resistance of a diffusive sample, thus making the choice
of the two-probe (or some other) geometry just the matter of
computational convenience.~\cite{mackinnon}
\begin{figure}
\centerline{\psfig{file=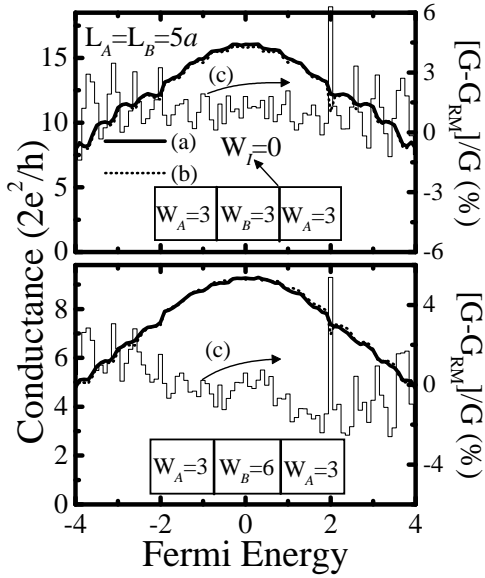,height=3.0in,angle=0} }
\vspace{0.2in} \caption{The disorder-averaged (over 200
configurations) conductance of a mesoscopic multilayer, composed
of strongly disordered interfaces and weakly disordered bulk
conductors, modeled on a lattice $17 \times 10 \times 10$. The
results are obtained from: (a) mesoscopic Kubo
formula~(\ref{eq:greenkubo}) evaluated for the whole multilayer;
and (b) following the resistor model Eq.~(\ref{eq:resistor}),
$G_{\rm RM}=(2R_A+R_B-2R_{\rm QPC})^{-1}$. The relative error
$[G-G_{\rm RM}]/G$ is plotted as (c).} \label{fig:layerw3noint}
\end{figure}

Since resistor model is expected to be retrieved in the limit of
completely diffusive scattering in the bulk and lack of phase
coherence,~\cite{schep2,bozec} in the quest for substantial
quantum effects induced deviations from this picture I start from
the opposite limit: a multilayer composed of clean (ballistic)
layers and disorder introduced on the interfaces, $W_A=W_B=0$ and
$W_I=30$. The disorder-averaged result is plotted in
Fig.~\ref{fig:mlbal2box} for two different thicknesses of the
interlayer. Even after disorder-averaging, the conductance
oscillates as a function of $E_F$. This clearly quantum effect is
a consequence of the size quantization caused by a coherent
interference of electrons reflected back and fort at the strongly
disordered interface. After adding the disorder into the outer
layers, the effect persists albeit with a smaller amplitude of
oscillations (Fig.~\ref{fig:mlbal3box}). However, while this is
plausible because of the interlayer being composed of only few
atomic monolayers~\cite{gijs} (i.e., its length is of the order
of $\lambda_F$), it is somewhat surprising that oscillations
increase with increasing the separation between the interfaces.
Furthermore, the oscillating conductance is still quite different
from a pure resonant tunneling conductance peaks which would
occur at the energies of bound states if interfaces are replaced
by the tunneling barriers~\cite{datta} (e.g., in our model this
can be generated by reducing the hopping integral $t_{\bf mn}$
between the outer layers and the interlayer~\cite{nikolic_qpc}).
Although it is obvious that this phenomenon cannot be accounted
by  the semiclassical resistor model, its answer is plotted for
the sake of comparison. Also, by adding impurities in the
interlayer we can follow the disappearance of conductance
oscillations (which is akin to the studies of disorder effects on
conductance quantization in ballistic conductors~\cite{szafer}).
The effect has almost vanished at $W=2$ although the mean free
path, e.g., around the band center, $\ell \approx 9$ (obtained
from Bloch-Boltzmann equation with Born approximation for the
scattering on a single impurity~\cite{nikolic_qrho}) is still
larger than $L$ (i.e., the transport within interlayer has not
yet reached the limit of fully diffusive bulk scattering).
\begin{figure}
\centerline{\psfig{file=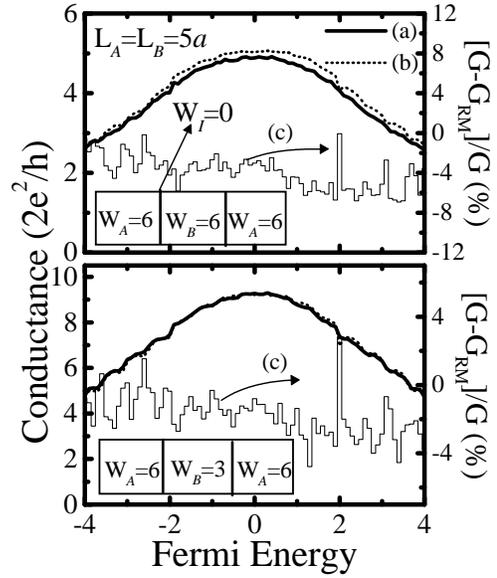,height=3.0in,angle=0} }
\vspace{0.2in} \caption{The same calculation as in
Fig.~\ref{fig:layerw3noint} (the same labels apply), but for a
slightly different multilayer with increased disorder strength in
the two outer layers.} \label{fig:layerw6noint}
\end{figure}

To sweep through the transition between fully quantum and
(expected) resistor model semiclassical description, a disorder is
introduced in both layers $A$ and $B$. The results are shown in
Figs.~\ref{fig:layerdifw3} and~\ref{fig:layerdif}. The disorder
$W=6$ is the strongest one in the Anderson model where one can
still use the semiclassical picture of transport at half-filling
(i.e., at $E_F=0$ we get~\cite{nikolic_qrho} $\ell \sim a$ for $W
\simeq 6$, but localization is postponed to much higher values
of~\cite{slevin} $W \simeq 16.5$). The other type of homogenous
layer is modeled with $W=3$, which is in the crossover between
the ballistic and the diffusive transport regime, since $\ell
\approx 4$ at $E_F=0$ and $L_A=L_B=5a$. The statistical error bars
on the disorder-averaged conductance (over $N_{\rm conf}$
samples), defined as $\Delta G = \sqrt{{\rm Var}\, G/N}$, are
smaller than the size of the dots (this further clarifies that
small conductances of the multilayers are not the consequence of
strong disorder, but are generated by combining few metallic
resistors and dirty interfaces in a series). The conductances of
individual layers are plotted in Fig.~\ref{fig:layerelem}. In the
first case $W_A=W_B=3$, the na\"{\i}ve application of the resistor
model, where $R_{\rm QPC}$ is neglected, leads to $G_{\rm
RM}^\prime$ ($G_{\rm RM}=1/R_{\rm RM}$ is the sum of component
resistances) being smaller than the quantum $G$ computed for the
multilayer as a single coherent unit. However, the subtraction of
four $R_{\rm QPC}$, which brings $G_{\rm RM}$ into the form of
Eq.~(\ref{eq:semilandauer}), gives conductance higher than
$G_{\rm RM} > G$. Here $R_{\rm QPC}$ is the resistance of a
ballistic conductor (``quantum point contact'') with a cross
section $A=100a^2$ (see Fig.~\ref{fig:layerelem}). According to
Eq.~(\ref{eq:semilandauer}), the resistances of homogeneous layer
components of the multilayer, which are plotted in
Fig.~\ref{fig:layerelem} (or Fig.~\ref{fig:interfaceg} for the
interface alone), are the sum of their ``intrinsic bulk''
resistance~\cite{mackinnon} and $R_{\rm QPC}$. Thus, when adding
these resistors the final sum should contain only one $R_{\rm
QPC}$ if resistor model is to be compared with the two-probe
resistance of the multilayer. In all other cases which include
layer with diagonal disorder $W=6$, both na\"{\i}ve and more
careful attempt (where $(N_R-1)$ contact resistances $R_{\rm QPC}$
are subtracted from the sum of resistance of component layers,
where $N_R$ is the number of bulk and interface resistances used
to obtain such sum) to use the resistor model give the
conductances which are greater than the ones computed for the
multilayer as a single conductor. The explanation which invokes
only extra scattering on the boundaries between different
components inside the multilayer, which does not work in the
first case $W_A=W_B=3$, is not the only one possible.

It was shown recently~\cite{nikolic_qrho} that even in weakly
disordered metals (where Boltzmann resistivity is practically
indistinguishable from the one computed from Kubo formula) clear
separation of a two-probe resistance into contact term and
diffusive bulk resistance, as implied by the standard arguments
of Eq.~(\ref{eq:semilandauer}), is tempered by localization
effects [i.e., terms of the order $\ell/L$ which are neglected in
Eq.~(\ref{eq:semilandauer})], and therefore possible only in very
weakly disordered systems. To elucidate the effects responsible
for the difference between the two ways of evaluating the
multilayer conductance in Figs.~\ref{fig:layerdifw3} and
\ref{fig:layerdif}, I apply the same analysis on a multilayer
where dirty interfaces are removed (i.e., $\varepsilon_{\bf m}=0$
on the sixth and twelfth plane along the $x-$axis). When $W_A \neq
W_B$ such multilayers are similar to the junction studied in
Sec.~\ref{sec:metaljunc} where interface is not explicitly
modeled as in Sec.~\ref{sec:interface}, but appears as a boundary
between two different homogeneous materials brought into a
contact. In the opposite case $W_A=W_B$, the sample akin to a
homogeneously disordered conductor whose conductance was studied
in Fig.~\ref{fig:homolead}. The results of this calculations,
plotted in Figs.~\ref{fig:layerw3noint} and
\ref{fig:layerw6noint}, hint that difference between resistor
model conductance $G_{\rm RM}$ and conductance $G$ of the
multilayer as one indivisible quantum-coherent unit is quite
minuscule, on the proviso that extraneous contact resistance terms
are properly subtracted. Even though the multilayers are
mesoscopic, where electrons retaining their phase coherence and
are subject thereby to quantum interference effects, disorder
averaging effectively destroys most of random interference terms
(a sample self-averages at finite temperatures when it can be
partitioned into a cubes of size $L_\phi$, as discussed
above).~\cite{altshuler} The surviving interference terms  are
exemplified by those generated by quantum interference effects on
special trajectories (i.e., Feynman paths), such as the closed
loops responsible for weak localization.~\cite{larkin} These
nonlocal weak localization effects, arising from the interference
between the amplitudes along the conjugated time-reversed loops
which span the whole phase-coherent sample, are responsible for
$G$ being slightly smaller than $G_{\rm RM}$ (localization
effects inside the individual resistors, summed in the resistor
model formula, are taken into account even by the semiclassical
approach). As expected in this picture, the discrepancy is very
small at $W=3$, and is increasing for stronger
disorder~\cite{nikolic_qrho} $W=6$. On the other hand,
conductance fluctuations sneak in through the fluctuations of
relative error $[G-G_{\rm RM}]/G$ of the resistor model as a
function of Fermi energy (cf. Figs.~\ref{fig:layerw3noint} and
~\ref{fig:layerw6noint}).

\section{Conclusions} \label{sec:conclusion}

I have studied different macroscopically inhomogeneous mesoscopic
disordered 3D conductors by employing both fully quantum
description, provided by the mesoscopic two-probe Kubo, and
semiclassical resistor model. Before embarking on the
technicalities of computations  some fundamental issues in the
quantum transport theory have been scrutinized: the relationship
between the Kubo formula in exact single-particle eigenstate
representation and Kubo formula for the finite-size sample
inspired by mesoscopic physics; the influence of the macroscopic
leads (``measuring apparatus'')  on the computed two-probe
conductance; and the transmission and transport properties of a
single dirty interface, which are markedly different from a
standard notions developed for bulk conductors. It was shown, by
exact computation on examples of homogenous disordered samples and
inhomogeneous metal junctions, that mesoscopic Kubo formula allows
one to obtain the reliable results for the static zero-temperature
conductance (which includes the properties of the attached leads)
characterizing noninteracting quasiparticle transport. On the
other hand, the evaluation of the traditional Kubo formula, based
on the exact diagonalization of the respective Hamiltonian, ends
up with a numerical discrepancy when compared to mesoscopic
methods applied to homogeneous samples (because of the necessity
to handle small numerical parameters, like the width of Lorentz
broadened delta function), but qualitative fails in the case of
inhomogeneous samples. This can be traced back to the conceptual
obstacles in applying methods derived for an infinite sample to a
finite-size system (even after it is extended through periodic
boundary conditions) where the problem of dissipation and effect
of the sample boundaries (which is crucial for the description of
mesoscopic devices) have been traditionally bypassed for the sake
of computational pragmatism.

The study of a single dirty interface, with specific Anderson
model type disorder, shows that its transmission properties are
well-accounted by the Schep-Bauer distribution, but these formula
applies approximately  also to thin slabs composed of a few of
such interfaces (which is important for experimental
investigations where one hardly deals with geometrical planes of
theoretical analysis). The nanoscale mesoscopic multilayers
containing such interfaces and ballistic bulk conductors exhibit
disorder-averaged oscillating conductance as a function of Fermi
energy. This effect of phase-coherence and quantum size effect is
slowly destroyed upon adding the disorder inside the layers.
However, even for diffusive scattering in the metallic layer
components, smooth disorder-averaged multilayer conductances can
not be completely accounted by the resistor model (which sums
layer and interface resistances as resistors connected in
series). Since this classical approach works well when interfaces
are removed (up to a tiny localization effects arising from
interference effects inside the whole sample), we can conclude
that just a single plane of a strongly disordered material is
enough to bring new quantum effects into the conductance of
mesoscopic metallic multilayered structures.

I am grateful to P. B. Allen for guidance and unlimited supply of
intriguing questions, and to  I. L. Aleiner and J. A. Verg\' es
for valuable discussions. This work was supported in part by NSF
grant no. DMR 9725037.

$*$ Present address: Department of Physics, Georgetown University,
Washington, DC 20057-0995.


\begin{references}
\bibitem{kubosubtle} Technical subtleties (like proper order of
non-commuting limits) in finding zero and nonzero surface terms
in the microscopic formulation of linear transport in open
systems are accounted in Ref.~\onlinecite{baranger-kubo}.
\bibitem{seemtobe} For a long time it seemed that these rigorous
(quantum) formulations of transport were merely serving to justify
the intuitively appealing Boltzmann approach. The new viewpoint
came with the first explicit calculation of quantum corrections
like weak localization~\cite{wl}---a quantum interference effect
which generates  a negative correction term to the Boltzmann
result, and is responsible at low temperatures for all of the
temperature and magnetic field dependence.
\bibitem{linearity} The relation of ``vanishing'' $eV$ to
the other relevant energy scales in the linear (quantum) transport
regime is reviewed in Ref.~\onlinecite{datta}.
\bibitem{andersonhub} Intricacies in the application of Kubo formula on
finite-size samples, ``extended'' through periodic boundary conditions,
were discovered also in some other condensed matter problems, e.g.
in the conduction in 1D Hubbard model, see Ref.~\onlinecite{andersonbook}.
\end{references}
\end{document}